\documentclass[
 reprint,
 amsmath,amssymb,
 aps,
 pra,
]{revtex4-2}

\usepackage{graphicx}
\usepackage{dcolumn}
\usepackage{bm}
\usepackage{hyperref}

\usepackage{bbm}
\usepackage{braket}
\usepackage{xcolor}

\newcommand{\beginsupplement}{%
	\setcounter{table}{0}
	\renewcommand{\thetable}{S\arabic{table}}%
	\setcounter{figure}{0}
	\renewcommand{\thefigure}{S\arabic{figure}}%
	\renewcommand{\theequation}{S\arabic{equation}}
}

\begin{document}

\title{Controlling Nonequilibrium Bose Condensation with Engineered Environments}

\author{Francesco Petiziol}
 \email{f.petiziol@tu-berlin.de}
 \affiliation{Technische Universit\"at Berlin, Institut f\"ur Theoretische Physik, Hardenbergstrasse 36, Berlin 10623, Germany}

\author{Andr{\'e} Eckardt}
 \email{eckardt@tu-berlin.de}
 \affiliation{Technische Universit\"at Berlin, Institut f\"ur Theoretische Physik, Hardenbergstrasse 36, Berlin 10623, Germany}

\date{\today}

\begin{abstract} 
Out of thermal equilibrium, bosonic quantum systems can Bose-condense away from the ground state, featuring a macroscopic occupation of an excited state or even of multiple states in the so-called Bose-selection scenario.
In previous work, a theory was developed that predicts, in which states a driven-dissipative ideal Bose gas condenses. Here, we address the inverse problem: Given a target state with desired condensate fractions in certain single-particle states, how can this configuration be achieved by tuning available control parameters? Which type of experimental setup allows for flexible condensation control? We solve these problems, on the one hand, by proposing a Bose `condenser', experimentally implementable in a superconducting circuit, where targeted Bose condensation into eigenstates of a chain of resonators is driven through the coupling to artificial quantum baths, realized via auxiliary two-level systems. On the other, we develop a theory to solve the inverse problem based on linear programming methods. We further discuss the engineering of transition points between different Bose condensation configurations, which may find application for amplification, heat-flow control, and the design of highly-structured quantum baths.
\end{abstract}

\maketitle

 Non-equilibrium Bose-Einstein condensation (BEC) has been widely explored in different platforms~\cite{Bloch2022}, such as photons in dye-filled cavities~\cite{Klaers2010, Schmitt2015, Walker2018, Hesten2018, Vorberg2018, Walker2019, Vlaho2021}, excitons~\cite{Butov2002, High2012, Alloing2014} and exciton-polaritons~\cite{Dang1998, Deng2002, Kasprzak2006, Balili2007, Deng2010, Byrnes2014, Barboux2016, Wertz2010} in (cavity) semiconductor heterostructures. BEC in these systems results from the interplay of thermalization with pump and loss, whose ratio determines the condensate mode. Non-equilibrium BEC of a different kind is predicted to occur, however, also at conserved particle number, when other mechanisms deprive the ground-state of its privileged role. Examples are open systems subject to time-periodic driving or to a strong competition between heating and cooling mechanisms~\cite{Vorberg2013, Vorberg2015, Schnell2017, Schnell2018, Schnell2023}. Steady states with multiple condensates in Bose-selected modes arise here from the nonequilibrium quantum-jump kinetics. Controlling their (fragmented) condensation pattern on demand is an appealing perspective for the design of, \textit{e.g.}, quantum signal amplifiers, multimode emitters, structured artificial quantum baths or heat-transport regulators. However, while a theory explaining Bose selection (BS) has been developed~\cite{Vorberg2013, Vorberg2015}, approaches to systematically engineer nonequilibrium BS, turning it into a practical resource, are missing. The key challenges are (i) how to realize controllable quantum-jump networks in realistic experimental conditions and (ii) how to solve the inverse problem of finding values of the control knobs yielding a target BS pattern. 
In this work, we solve both problems by proposing a concrete experimental setup where nonequilibrium BEC of photons can be controlled in a superconducting circuit using synthetic reservoirs and by developing methods to reverse-engineer control parameters yielding desired condensation patterns. We demonstrate the success of this procedure in simulations with realistic values of experimental parameters and propose an application in the design of a quantum switch for heat transport.  

\begin{figure}
\includegraphics[width=\linewidth]{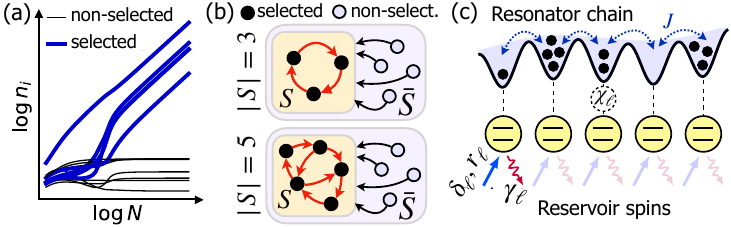}
\caption{(a) Sketch of the BS effect: selected states acquire a macroscopic occupation $\propto N$, while the occupation of non-selected states saturates for large $N$. (b) Network representation of examples of asymmetry matrices $A$ compatible with BS (proof in the SM~\cite{SM}). (b) Sketch of a `Bose condenser': a resonator chain (blue-colored), where BEC occurs in selected eigenmodes, is dispersively coupled to driven-damped reservoir two-level systems (yellow).} 
\label{fig:Figure1}
\end{figure}

{\it Bose selection} can occur in a system of $N$ non-interacting bosons exchanging energy with its environment, whose dissipative dynamics is described by a many-body Lindblad master equation ($\hbar=1$), 
\begin{equation} \label{eq:mbme}
\dot{\hat{\rho}}_S = -i[\hat H_S,\hat{\rho}_S]+\sum_{ij} R_{ij} D[\hat{L}_{ij}]\hat{\rho}_S\ ,
\end{equation}
 for the density operator $\hat{\rho}_S$. Given single-particle eigenstates $\ket{i}$ with energy $E_i$, Eq.~\eqref{eq:mbme} involves the Hamiltonian $\hat H_S=\sum_iE_i\hat{n}_i$, dissipators $D[\hat{L}_{ij}]\hat{\rho}_S=\hat{L}_{ij}\hat\rho_S\hat{L}_{ij}^\dag -\{\hat{L}_{ij}^\dag\hat{L}_{ij},\hat{\rho}_S\}/2$ with rates $R_{ij}$ for quantum jumps from $\ket{j}$ to $\ket{i}$, jump operators $\hat{L}_{ij}=\hat{c}^\dagger_i\hat{c}_j$, and annihilation and number operators $\hat{c}_i$ and $\hat{n}_i=\hat{c}^\dagger_i\hat{c}_i$ for a boson in the $i$th single-particle state. The rates are assumed to realize a fully-connected network, implying a unique steady state~\cite{Vorberg2013}, which becomes the equilibrium state, only when $R_{ji}/R_{ij}=\exp[-\beta(E_i - E_j)]$, with inverse temperature $\beta$. 
Although the bosons are non-interacting, the coupling to the bath(s) makes the problem interacting, as the dissipator is quartic in the $\hat{c}_i$. The hierarchy of Eqs.~for the $m$-point correlators $\mathrm{tr}[\hat{\rho}_S \hat{n}_{i_1} \hat{n}_{i_2}\dots\hat{n}_{i_m}]$, resulting from this interacting problem, can be truncated through a mean-field approximation, which yields nonlinear kinetic Eqs.~of motion for the $n_i=\mathrm{tr}[\hat{\rho}_S \hat n_i ]$, namely $\dot{n}_i = \sum_{j} \big[R_{ij} {n}_j (1+{n}_i) - R_{ji} {n}_i(1+{n}_j)\big]$ ~\cite{Vorberg2013, Vorberg2015} (see the SM~\cite{SM} for a brief review). The nonlinearity is related to Bose statistics, giving rise to a dependence of the many-body rate on the occupation a particle jumps to, known as bosonic enhancement (or stimulated emission). It is responsible for the emergence of BS in the steady state, satisfying $\dot{n}_i=0$, where a single or multiple selected states acquire a macroscopic occupation $\propto N$ in the large-$N$ limit, while the occupation of all other states saturates [sketch in Fig.~\ref{fig:Figure1}(a)]. The set of selected states $S$ (and its complement $\bar{S}$) can be predicted by considering a large-$N$ expansion $n_i=\nu_i N +\sum_{\alpha = 1}^\infty\nu_i^{(\alpha)}/N^{\alpha-1}$. It reveals that condensation is ruled, in leading order, by the rate-asymmetry matrix $A_{ij}=R_{ij}-R_{ji}$ through the set of (in)equalities~\cite{Vorberg2013, Vorberg2015}
\begin{equation} \label{eq:condcond}
 A_{\bar{S} S} \bm{\nu}_{S} < 0, \quad  A_{SS} \bm{\nu}_S = 0,\quad \bm{\nu}_{\bar{S}}=0, \quad \bm{\nu}_S>0 ,
\end{equation}
which ensure that $n_i\ge0$ for all $i$. The inequalities are understood elementwise and $\nu_X$ and $A_{XY}$ denote subvectors and matrix blocks, respectively, with $X,Y \in \{S,\bar{S}\}$.
 
By depicting $A$ as a network with edges pointing from the $j$th to the $i$th node if $R_{ij}>R_{ji}$, as in Fig.~\ref{fig:Figure1}(b), physical intuition about rates $R_{ij}$ admitting BS can be gained. The first condition in~\eqref{eq:condcond} is satisfied, \textit{e.g.}, if all non-selected states directly feed selected states, pointing at them in the network. Selection in a single state $\ket{c}$ occurs iff $R_{cj}>R_{jc}$ $\forall j$~\cite{Vorberg2013}, thus making $\ket{c}$ `ground-state-like'. For multimode BS, no state in $S$ must be a global attractor, as imposed by the second condition, demanding $A_{SS}$ to have a non-trivial kernel vector $\nu_{S,i}>0$. For three-state selection, this is possible only for a loop-like configuration as in Fig.~\ref{fig:Figure1}(b)-top~\cite{Knebel2015, SM}. For larger sets $S$, more complex network topologies and rate imbalances are needed, with an example for $|S|=5$ in Fig.~\ref{fig:Figure1}(b)-bottom, and the resulting condensate fractions depend nonlinearly on $A_{SS}$~\cite{Knebel2013, Knebel2015} (and not only on the topology of the directed network). This highlights the difficulty in engineering specific condensation patterns, seemingly requiring one to assemble intricate rate-networks edge by edge, which is beyond experimental reach. We will, therefore, consider the more realistic scenario, where the rates depend on a number of control parameters, and show how these parameters can be optimized for achieving the desired BS pattern.

{\it A superconducting ``Bose condenser''.} 
We propose an experimental implementation given by an array of $M$ microwave resonators in a superconducting circuit~\cite{Krantz2019, Blais2021, Roushan2017, Yan2019, Ma2019, Carusotto2020, Marcos2012}. Each resonator is dispersively coupled to an ancillary transmon qubit subject to coherent driving and loss, which implements a narrow-band artificial bath (hereafter described by a spin 1/2) [Fig.~\ref{fig:Figure1}(c)]. Owing to the harmonicity of the resonator spectrum, the array hosts noninteracting microwave photons~\cite{Carusotto2020}, whose condensation we aim to control. The dynamics is described by the master equation 
\begin{equation} \label{eq:resspinme}
\frac{d\hat{\rho}}{dt} = -i [\hat{H}_S + \hat{H}_B +\hat{H}_{SB}, \hat{\rho}] + \sum_{j} \gamma_\ell D[\hat{\sigma}_\ell^-]\hat{\rho} \ ,
\end{equation}
for the density matrix $\hat{\rho}$ of the combined resonators-spins system, where $\gamma_\ell$ is the decay rate of the $\ell$th spin and $\hat{\sigma}_\ell^\alpha$ ($\alpha=x,y,z,\pm$) are Pauli matrices. The Hamiltonian $\hat{H}_S$ of the bosonic system is given by $\hat{H}_S = \sum_{\ell} \omega_\ell \hat{a}_\ell^\dagger \hat{a}_\ell - \sum_{\ell,\ell'} J_{\ell\ell'}\hat{a}_{\ell}^\dagger \hat{a}_{\ell'}=\sum_i E_i\hat{c}_i^\dagger \hat{c}_i,$ where $\hat{a}_\ell$ and $\hat{a}_\ell^\dagger$ are the annihilation and creation operators of a photon in the $\ell$th resonator, which has transition frequency $\omega_\ell$, and $J_{\ell\ell'}>0$ are the tunnelling strengths. We choose the example of $\hat{H}_S$ describing a 1D chain, with $J_{\ell\ell'} = J$ for nearest neighbours $\ell$ and $\ell'$. We include a weak disorder, $|\omega_\ell-\omega_{\ell'}| \sim J_{\ell\ell'}$, to break symmetries inducing multiple identical level spacings in the spectrum. Concretely, we use $\omega_\ell/J = \ell/10 + \varepsilon_\ell$, where $\varepsilon_\ell$ are random numbers uniformly distributed in $[0,1)$~\cite{SM}, though the specific values are not important for the target physics. These values of $\omega_\ell$ are also chosen, such that the eigenmodes $\ket{i}$ are still delocalized over several lattice sites for the system-sizes considered. These features will be useful for realizing efficient dissipation engineering. The artificial-bath Hamiltonian $\hat{H}_B=\sum_{\ell} \delta_{\ell} \left(\hat{\sigma}^z_\ell + r_{\ell} \hat{\sigma}^x_\ell\right)/2$ describes the transmon spins (in a frame rotating at the driving frequency), where $\delta_{\ell}$ is the detuning of the drive and $r_{\ell}$ is the ratio between its Rabi frequency and $\delta_\ell$. In the dispersive-coupling regime, the spins are far detuned from the resonators. Resonant exchange of excitations is thus strongly unfavored, implying photon number conservation as desired (weak particle loss is discussed later). Other system--artificial-bath couplings as used, \textit{e.g.}, in Refs.~\cite{Ma2019, Mi2024} would involve particle exchange instead. Moreover, in the dispersive regime, undesired photon-photon interactions induced by the coupling to the spins can also be neglected. The resonator-spin coupling is described by the Hamiltonian $\hat{H}_{SB} = \sum_{\ell} \chi_{\ell} \hat{a}^\dagger_{\ell} \hat{a}_{\ell} \hat{\sigma}_{\ell}^z$, provided the mean number of photons in a resonator does not exceed a critical value~\cite{Blais2004, Blais2021}. For the parameters used below, tens to hundreds of photons per site are allowed, sufficient for BS. Condensation control will be achieved through the coherent drive and by tuning the value of the dispersive coupling $\chi_\ell$, in order to leverage the impact of each artificial bath on the quantum-jump network. Control of $\chi_\ell$ is implemented, for instance, with frequency-tuneable transmons, by adapting their detuning from the resonators~\cite{Koch2007, Swiadek2023}, or ones with tuneable coupling~\cite{Chen2014, Roushan2017b}. Note that the use of nonlinear (two-level) elements as artificial bath, rather than additional cavity modes~\cite{Murch2012, Hacohen2015, Kapit2017, Petiziol2022}, is crucial: the dispersive coupling to the latter would only give a mutual state-independent energy shift~\cite{Blais2021}, rather than the density-density-like coupling in $\hat{H}_{SB}$. 

\textit{Engineered quantum-jump rates.} Although $\hat{H}_{SB}$ commutes with the boson number $\hat{a}_\ell^\dagger\hat{a}_\ell$ in a single resonator, preventing particle loss, it has sizeable matrix elements between the non-site-local modes $\ket{i}$, giving rise to non-zero off-diagonal elements of $R_{ij}$ tuneable via the spins. Intuitively, this enables for dissipation engineering with the following mechanism. The coherent-control parameters $\delta_\ell$ and $r_\ell$ are adapted such that the spin's excitation energy, $\mathcal{E}_\ell = \delta_\ell \sqrt{1 + r_\ell^2}$, matches the energy spacing between states $\ket{i}$ and $\ket{j}$ in the array. The coupling drives a coherent excitation exchange involving a $\ket{i}\leftrightarrow\ket{j}$ transition, with a matrix element $\chi_{ij}^{(\ell)} = r_\ell \chi_\ell M_{ij}^{(\ell)}/\sqrt{1+r_\ell^2}$, where $M_{ij}^{(\ell)}= \bra{i}\hat{a}_\ell^\dagger \hat{a}_\ell \ket{j}$, which we shall compute below. This excitation flip-flop is interrupted by the strong spin damping, which drags the spin to its drive-dependent steady state and results in dissipative quantum jumps in the bosonic chain at rates $R_{ij}$ and $R_{ji}$. Here the spin relaxation time plays a role similar to a short bath correlation time in the more conventional Born-Markov scenario with a large bath. 
 
To derive the rates $R_{ij}=\sum_\ell R_{ij}^{(\ell)}$ of Eq.~\eqref{eq:mbme} from Eq.~\eqref{eq:resspinme}, consider first a single reservoir-spin coupled to the $\ell$-th resonator (further technical details on the derivation are given in the SM~\cite{SM}). The generalization to multiple reservoirs is straightforward, as their rates simply sum up. Representing Eq.~\eqref{eq:resspinme} in a diagonal basis for the spin, terms corresponding to non-secular (off-diagonal) elements of the Kossakowski matrix of the dissipator are off-resonant and can be neglected in rotating-wave approximation (RWA). This is justified provided the spin level splitting $|\mathcal{E}_\ell|$ is much larger than its decay rate $\gamma_\ell$, namely $|\mathcal{E}_\ell| \gg \gamma_\ell$ (\textit{Approximation I}). Equation~\eqref{eq:resspinme} becomes
\begin{multline}
\dot{\hat{\rho}} = -i \left[\sum_{i} E_i\hat{n}_i -\sum_{i,j} \chi_{ij}^{(\ell)}\hat{L}_{ij} \left(\frac{\hat{\sigma}_\ell^z}{r_\ell} + \hat{\sigma}_\ell^x\right)
 -  \frac{\mathcal{E}_\ell}{2} \hat{\sigma}_\ell^z, \hat\rho \right] \\
  +  \gamma_\ell^+ D[\hat{\sigma}_\ell^+]\hat{\rho} + \gamma_\ell^- D[\hat{\sigma}_\ell^-]\hat{\rho} + \gamma_\ell^z D[\hat{\sigma}_\ell^z]\hat{\rho}\ , \label{eq:anotherme}
\end{multline}
including $\mathcal{E}_\ell =\delta_\ell \sqrt{1 + r_\ell^2}$ and the dressed decay rates $ \gamma_\ell^+ = \gamma_\ell \cos(\theta_\ell)^4,\ \gamma_\ell^- = \gamma_\ell \sin(\theta_\ell)^4, \ \gamma_\ell^z = \gamma_\ell r_\ell^2/[4(1 + r_\ell^2)]$, where $\theta_\ell= \arctan(r_\ell)/2$ is the spin's mixing angle.

Considering the case in which $\mathcal{E}_\ell$ is (quasi)resonant with the level spacing $E_{ij}=E_i-E_j>0$ in the system, the interaction terms $\hat{L}_{ij}\hat{\sigma}_\ell^+$ and $\hat{L}_{ij}^\dagger\hat{\sigma}_\ell^-$ become resonant. Other interaction terms can be neglected, in RWA, if $E_{ij}$ is much larger than the effective coupling $\chi_{ij}^{(\ell)}$, $E_{ij} \gg |\chi_{ij}^{(\ell)}|$ (\textit{Approximation II}). Next, we trace out the spin, treating it as an environment degree of freedom, with $\chi_{ij}^{(\ell)}$ representing a system-bath coupling, and following standard Born-Markov-secular derivations~\cite{Breuer2007}. This approach is justified, despite the spin being far from constituting a true bath, provided the spin relaxes much faster than the timescale of interaction with the system, $|\chi_{ij}^{(\ell)}| \ll \gamma_\ell$ (\textit{Approximation III}). Then, the spin state is negligibly affected by this interaction and can be approximated as constant. Moreover, the state of the system and the artificial bath can be approximated as factorized. The dynamics of the system is then described by the Markovian master equation~\cite{Breuer2007}, $\dot{\hat{\rho}}_S = \int_0^{+\infty} ds \ \mathrm{tr}_\sigma\big[ \hat{H}(t), [\hat{H}(t-s), \hat{\rho}_S\otimes \hat{\rho}_\sigma]\big],$ where $\mathrm{tr}_\sigma$ and $\hat{\rho}_\sigma$ denote the trace over the spin degrees of freedom and the spin's steady state, respectively. $\hat{H}(t)$ represents the Hamiltonian of Eq.~\eqref{eq:anotherme} in the interaction picture. 

Following standard manipulations~\cite{Breuer2007}, the Markovian master Eq.~can be brought into the form of Eq.~\eqref{eq:mbme}, within approximations I-III, with rates
$ R^{(\ell)}_{ij}= |\chi_{ij}^{(\ell)}|^2 \mathcal{S}_{\ell}^+(-E_{ij})$ and $R^{(\ell)}_{ji}= |\chi_{ij}^{(\ell)}|^2 \mathcal{S}_\ell^-(E_{ij})$ for $\delta_\ell>0$. The quantum noise spectra $\mathcal{S}_\ell^{\pm}(\omega)  = \int_{-\infty}^{\infty}dt \ e^{i\omega t} \mathrm{tr} \{[\hat{\sigma}_\ell^\pm(t)]^\dagger \hat{\sigma}_\ell^\pm(0) \hat{\rho}_{\sigma}\}$ are computed from the spin's steady state $\hat{\rho}_{\sigma}$ and the correlation functions~\footnote{For negative detuning, the derivation of the master equation is the same as above, only with the role of $\hat{\sigma}_\ell^-$ and $\hat{\sigma}_\ell^+$ [and thus $\mathcal{S}_\ell^-(\omega)$ and $\mathcal{S}_\ell^+(\omega)$] exchanged}. Here, the operators $\hat{\sigma}_\ell^\pm(t)$ are interaction-picture representations of $\hat{\sigma}_\ell^\pm$. 
 The correlation functions are computed in the SM~\cite{SM} from the optical Bloch equations for the spin and the quantum regression theorem~\cite{Carmichael1999, Breuer2007}. We obtain the quantum noise spectra
\begin{equation} \label{eq:spectrum_pm}
\mathcal{S}_\ell^\pm(\omega) = \frac{\gamma_\ell^\mp }{\gamma_\ell^+ + \gamma_\ell^-}\frac{2 \Gamma_\ell}{(\omega \pm \mathcal{E}_\ell)^2 + \Gamma_\ell^2}\ ,
\end{equation}
characterized by a linewidth $ \Gamma_\ell = (\gamma_\ell/4)\left[3 - 1/(1+r_\ell^2)\right]$. For a positive $E_{ij}>0$ and detuning $\delta_\ell>0$, the rates can be rewritten as 
\begin{equation} \label{eq:rates_t}
R^{(\ell)}_{ij}= \gamma_{\ell}^- |\chi_{ij}^{(\ell)}|^2  \mathcal{S}_{\ell}(E_{ij}), \ 
 R^{(\ell)}_{ji}= \gamma_{\ell}^{+}  |\chi_{ij}^{(\ell)}|^2 \mathcal{S}_{\ell}(E_{ij}),
\end{equation}
with $\mathcal{S}_\ell(\omega)=2\Gamma_\ell/[(\omega - |\mathcal{E}_\ell|)^2 + \Gamma_\ell^2]$. For $\delta_\ell<0$, the rates have the same form~\eqref{eq:rates_t}, but with $\gamma_{\ell}^\pm$ exchanged. The driven spin dynamics determines the quantum-jump ratio $R^{(\ell)}_{ij}/R^{(\ell)}_{ji}=\gamma_\ell^{-}/\gamma_\ell^{+}=\tan(\theta_\ell)^4$ via the mixing angle $\theta_\ell$. For $\delta_\ell>0$ and $\theta_\ell\ll 1$, the spin dissipator in Eq.~\ref{eq:anotherme} drags the spin to its ground state, favouring the processes $\hat{L}_{ij}\hat{\sigma}^+_\ell$ which increase the system energy, but blocking the inverse processes $\hat{L}_{ij}^\dagger\hat{\sigma}^-_\ell$ which lower the energy. The opposite occurs for $\theta_\ell\approx \pi/2$. Intermediate values of $\theta_\ell$ allow both processes with a finite rate and are optimal to ensure a sizeable matrix element $\chi_{ij}^{(\ell)}\propto r_\ell$, while complying with the weak-driving regime desired in experiments. Summarizing, each artificial bath can be tuned to increase or decrease the system energy with a controllable rate ratio and the corresponding rates are enhanced around the peak of $\mathcal{S}_\ell(\omega)$. 
Alternative control scenarios based on a Floquet modulation of the bosonic system may be possible, but would make the control more difficult. Indeed, strong driving alters the shape of the modes $\ket{i}$, which become Floquet states, and gives rise also to side-band quantum jumps (see, \textit{e.g.}, Refs.~\cite{Vorberg2013, Schnell2023}). 

The validity of Eq.~\eqref{eq:rates_t} requires the hierarchy $E_{ij}\gg \gamma_\ell \gg |\chi_{ij}^{(\ell)}|$ between the system energy gap $E_{ij}$, the spin decay rates $\gamma_\ell$ and the transition matrix elements $\chi_{ij}^{(\ell)}$, resulting from Approximations I-III. This hierarchy defines the operating regime of the proposed Bose condenser, which is easily met in state-of-the-art superconducting devices~\cite{Ma2019}. In particular, we envision implementations involving a number of resonators of the order of ten, each coupled to a superconducting qubit, which are close to the setup of recent experiments~\cite{Ma2019, Yan2019} and suffice for the potential applications described below. Given that system gaps are of the order of the tunnelling strength $J$ for such system sizes, we may choose a value $J/2\pi \sim 30$ MHz and decay rates $\gamma_\ell\sim 1-10$ MHz (realized, \textit{e.g.}, in~\cite{Ma2019}). The Rabi frequencies $|\delta_\ell |r_\ell$ used in our examples are lower than $1.5J\sim2\pi\times 40$ MHz in value, thus meeting standard constraints for microwave drives on transmons. We then obtain $|\chi_{ij}^{(\ell)}| \sim 10^{-1} J$ by also restricting $\chi_j$ to a maximal value $\chi_{\rm max}\sim 0.15J\sim2\pi\times 4.5$ MHz~\cite{Krantz2019}. In turn, this yields engineered quantum-jump rates $R_{ij}\sim10^{-2}J$, much stronger than typical photon loss rates for microwave resonators in circuit QED~\cite{Blais2021}. Even in case losses are not compensated for (as we propose below), the system can thus form a Bose-selected steady-state well before serious particle loss occurs. 

\begin{figure}[t]
\includegraphics[width=\linewidth]{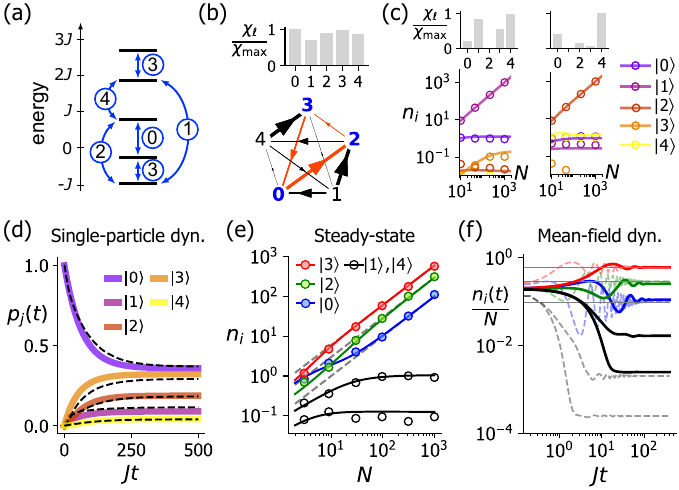}
\caption{(a) Single-particle spectrum of the five-site chain. 
The arrow numbers indicate the reservoir-spin used to induce the dissipative transition between the corresponding linked states. (b), (d), (e), (f) show the controlled BS in states $\ket{0}$, $\ket{2}$, $\ket{3}$, namely the values of the control parameters $\chi_\ell$ and the resulting asymmetry-matrix network [(b)-- the arrows and their thickness represent $\mathrm{sign}(A_{ij})$ and $|A_{ij}|$, respectively], the single-particle dynamics [(d)], the steady state as a function of the total particle number $N$ [(e)] and the mean-field dynamics for $N=50$ (solid) and $N=500$ (dashed) [(f)]. Symbols and line types are explained in the main text. (c) Controlled BEC into single modes $\ket{1}$ and $\ket{2}$, and corresponding values of $\chi_\ell$.}
\label{fig:2}
\end{figure}

{\it Programmable BECs.} To convert BS into a control problem in the Bose condenser, we identify controllable coefficients $\bm{z}$ leveraging the different bath contributions as $z_\ell=\mathrm{sign}(\delta_\ell)(\chi_\ell/\chi_{\rm max})^2$, leading to a decomposition of the total asymmetry matrix as $A= \sum_{\ell} z_{\ell} A^{(\ell)}$. 
We then construct a recipe to reverse-engineer a $\bm{z}$ giving a target BS pattern $\bm{\nu}$. From the conditions~\eqref{eq:condcond} we derive a new set of inequalities for $\bm{z}$,
\begin{equation} 
\quad (B \bm{z})_{\bar{S}} < 0, \quad (B \bm{z})_S = 0\ ,
\label{eq:cond_on_z}
\end{equation}
with $ B_{ij}\equiv\sum_{\ell\in S} A_{i\ell}^{(j)} \nu_{\ell}$. If a solution $\bm{z}$ exists, it will stabilize the targeted steady state with $\bm{\nu}$ by construction. To efficiently search for solutions of Eq.~\eqref{eq:cond_on_z}, we rephrase the inequalities as constraints in a linear program and solve them using linear-programming routines~\cite{SM, Knebel2015}. This procedure represents a powerful framework to reverse-engineer BS patterns. Let us exemplify it by considering a five-site chain, possessing the single-particle spectrum shown in Fig.~\ref{fig:2}(a). The specific values of the system parameters are not crucial for the algorithm proposed to return a successful condensation protocol and the values used in the following, specified in the SM~\cite{SM}, are within the experimentally accessible ranges discussed above. The energies of the reservoir spins are set in resonance with different energy-level distances in the system, such that the peaks of their spectral density approximate the overall connectivity sketched in Fig.~\ref{fig:2}(a), ensuring that every state is reachable via quantum jumps. The energy-spacing-to-spin association is chosen by verifying numerically that strong matrix elements $|M^{(\ell)}_{ij}|$ are attained. 

We solve Eq.~\eqref{eq:cond_on_z} for the control variables $\bm{z}$ by targeting BS into a chosen set of three modes, $\{\ket{0}, \ket{2}, \ket{3}\}$, with target condensate fractions $[\bm{\nu}_{S,0}, \bm{\nu}_{S,2}, \bm{\nu}_{S,3}]=[1/10, 3/10, 6/10]$, finding the control values $\chi_\ell$ and the asymmetry matrix depicted in Fig.~\ref{fig:2}(b). The latter features the loop structure of Fig.~\ref{fig:Figure1}(b)-top within the set of selected states (red arrows), needed to sustain BS. To verify the validity of the effective rates~\eqref{eq:rates_t} derived, we compare in Fig.~\ref{fig:2}(d) the single-particle dynamics given by those rates with the master Eq.~\eqref{eq:resspinme}, confirming good agreement. The achievement of the desired BS pattern in the steady state is shown in Fig.~\ref{fig:2}(e), where saturation of non-selected states starts for $N\gtrsim30$ and the target fractions are reproduced faithfully already for $N=100$. The steady-state occupations are computed numerically here with different methods corresponding to the approximation layers used in designing the control protocol (recapitulated in~\cite{SM}), to confirm their reliability: solution of the asymptotic large-$N$ theory of Eq.~\eqref{eq:condcond} underpinning the algorithm~\eqref{eq:cond_on_z} via linear programming~\cite{Knebel2015} (dashed lines), long-time propagation of the mean-field kinetic Eqs.~(solid lines), and long-time propagation of the populations of $\hat{\rho}_S(t)$ from Eq.~\eqref{eq:mbme} via a quasi-exact Monte-Carlo-quantum-trajectory-type unravelling~\cite{Vorberg2015} (bullets, averaged over $10^3$ trajectories). Thanks to the bosonic enhancement in the many-body rates, the evolution converges rapidly to its BS steady state [Fig.~\ref{fig:2}(f)]. The depletion of non-selected states takes place within few tens of tunnelling times for $N=50$ (solid lines), and faster and faster as $N$ increases [$N=500$ in Fig.~\ref{fig:2}(f)---dashed lines]. By exploring different connectivities through different energy-spacing--to--spin associations and solving for $\bm{z}$, protocols giving selection into any triplet of states, with the same occupation fractions as above, are also found~\cite{SM}. Condensation in individual states can also be achieved, see Fig.~\ref{fig:2}(c) showing examples with $\ket{1}$ and $\ket{2}$. 

\begin{figure}[t]
\includegraphics[width=\linewidth]{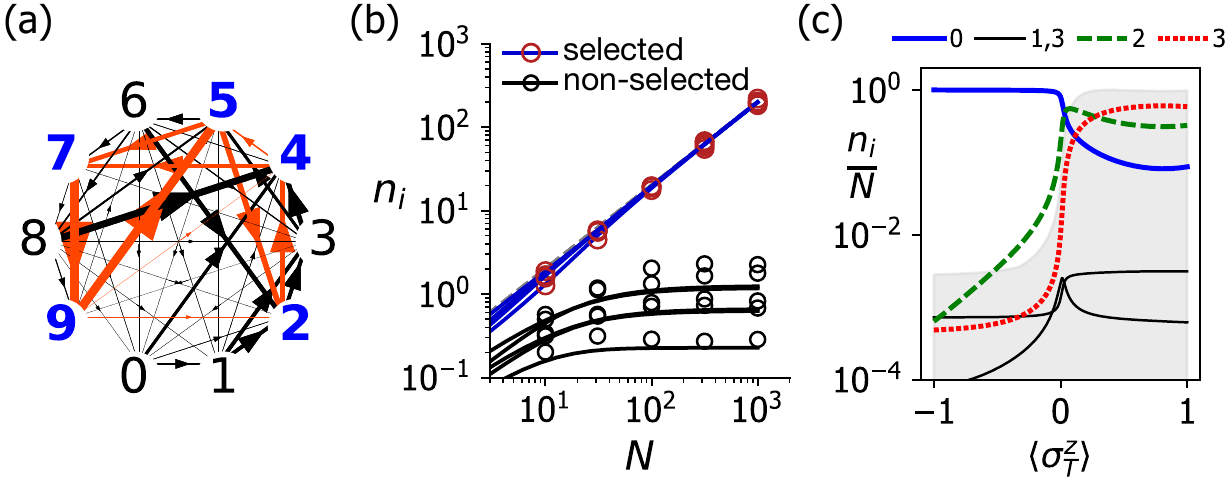}
\caption{(a) Asymmetry matrix for selection of five modes (blue) in a ten-site chain; transitions within $S$ are colored red. (b) Achievement of BS with requested equal occupancy (solid lines: mean-field theory, circles: quasi-exact average of 500 quantum trajectories. (a) BEC pattern vs.\ expectation value $\langle \hat{\sigma}_T^z\rangle$ of the `trigger' spin for $N=10^2$. The grey area indicates the total value of the heat current $-\sum_{\ell\in B_X} j_h^{(\ell)}$ induced in the energy-extracting artificial baths (denoted as the set $B_X$), normalized by its maximum.}
\label{fig:10s}
\end{figure}

To exemplify controlled selection into more than three states, we target five modes of a ten-site chain. The reservoir spins realize rate asymmetries $A_{ij}$ for which each target mode is strongly connected to at least two other selected states. Choosing states $\{\ket{2},\ket{4},\ket{5},\ket{7},\ket{9}\}$ with equal occupancy and solving Eqs.~\eqref{eq:cond_on_z} for $\bm{z}$ gives the asymmetry matrix of Fig.~\ref{fig:10s}(a), where we find a rate structure of the type of Fig.~\ref{fig:Figure1}(b)-bottom within $S$. The target BS pattern is successfully attained [Fig.~\ref{fig:10s}(b)]. The examples considered represent ideal system sizes for potential implementation: they are sufficient to demonstrate the desired effect, while being close to setups already realized~\cite{Ma2019, Yan2019}. Also, in view of potential applications as they are discussed below, larger system sizes do not provide an advantage. For much larger chain length, we expect that inducing fully-arbitrary condensation patterns will become increasingly challenging. As the spectrum becomes denser for increasing length, the transitions induced by the artificial baths will become less selective. Moreover, the number of level spacings grows much faster than the number of control parameters, assuming the use of one reservoir-spin per resonator. Still, the control method proposed here may be used to drive BS into states within a certain energy window of finite width, rather than sharply in individual states.

We assumed until now negligible particle loss, namely that the resonators' relaxation is much slower than the engineered dissipation. In the SM~\cite{SM}, we analyze the impact of weak losses and the use of spin reservoirs realizing particle pumps to counteract them. If the total mean particle number is kept large, BS is to good approximation solely dictated by the matrix $A$, rather than by pump and loss. It is, thus, still successfully controlled with the above methods. This is easily explained by the fact that BS derives from terms in the mean-field equation that are quadratic in $n_i$, whereas pump and loss enter linearly. For the same reason, the BS steady state is attained well before substantial losses occur. Pump and loss may also be used as additional control parameters.

{\it A condensate switch.} Realizing nonequilibrium phases which are {\it per se} robust, but tuned to be sensitive to few selected parameters, is a promising resource for, \textit{e.g.}, amplification and sensing. We can use the ability to control BS to design similar conditions in the Bose condenser. The system can be tuned to a transition point between two (or more) BS configurations, such that the state of an additional ``trigger'' system coupled to it determines which condensation pattern arises. For instance, in the five-site device of Fig.~\ref{fig:2}(a), the three-state BS turns into ground-state condensation, if the quantum jump governed by the third reservoir-spin is forced towards the ground state. Consider then a coupling of the spin's detuning $\delta_3$ to an additional trigger spin (`$T$'), through the Hamiltonian $\hat{H} = \delta_3 (r_3 \hat{\sigma}_3^x + \hat{\sigma}_3^z \hat{\sigma}_T^z) + \omega_T\hat{\sigma}_T^z/2$. Enforcing $|\mathcal{E}_3| \approx |E_{20}|$, the bosonic steady state then depends on the state of $`T'$, through the renormalization of $\delta_3$ by $\langle \hat{\sigma}_T^z \rangle$. Depending on whether $\langle \hat{\sigma}_T^z \rangle = \pm 1$, the chain will exhibit either three-state or ground-state selection, as numerically verified in Fig.~\ref{fig:10s}(c). A similar device may be used, for instance, as a switch that activates or deactivates the transport of macroscopic energy currents through the system. Indeed, the heat current $j_h^{(\ell)}$ generated in the system by the coupling to the $\ell$th artificial bath is given, within the mean-field theory, by $j_h^{(\ell)}=\mathrm{tr}[\dot{\hat{\rho}}_S \hat{H}_S] \approx \sum_{ij} (E_j-E_i) n_i(n_j+1)R_{ji}^{(\ell)}$~\cite{Vorberg2013}. The current is bosonically enhanced (quadratic in the $n_i$) for transitions between selected states, while it is only linearly enhanced for transitions between a selected and a non-selected state. When multiple condensates are present, the system has thus the ability to absorb and emit much more energy at large particle numbers. This feature may be of particular interest for the use of the BS system as a junction ruling quantum heat transport in a more complex device, or as an artificial bath with a highly-structured spectral density. The heat-current enhancement is shown in Fig.~\ref{fig:10s}(c) for the five-site Bose condenser (grey shading), as the system passes from a single-condensate to a three-condensate phase. A interesting perspective in this context is also the classification of such nonequilibrium phase transitions among BS patterns. The latter may be distinguished in terms of the topology of the quantum-jump network, for instance exploiting the mathematical similarity between the BS asymmetry-matrix theory and Lotka-Volterra systems in evolutionary game theory~\cite{Geiger2018}.

{\it Conclusion.} We showed how non-equilibrium BEC can be controlled via the coupling to engineered quantum baths. We designed a physical setup, implementable, \textit{e.g.}, in superconducting-circuit architectures, granting a handle on the condensate location and fragmentation, which can thus be shaped on demand. These results pave the way to applications in the control of heat transport, amplification, and quantum bath engineering. Combined with topologically non-trivial band structures, controlled selection may facilitate edge-mode detection~\cite{Furukawa2015}, or realize topological-laser-like~\cite{Bandres2018, Harari2018} steady states.

\acknowledgments{This research was funded by the Deutsche Forschungsgemeinschaft (DFG, German Research Foundation) via the Research Unit FOR 2414 - project No. 277974659.} 

\bibliography{biblio}

\cleardoublepage

\beginsupplement
\setcounter{page}{1}
\onecolumngrid

\begin{center}
{\bf \large Supplemental Material to} \\
\vspace{0.2cm}
{\bf \large Controlling Nonequilibrium Bose Condensation with Engineered Environments} \\
\vspace{0.4cm}

{ Francesco Petiziol and Andr\'e Eckardt}\\

{\itshape
Technische Universit\"at Berlin, Institut f\"ur Theoretische Physik, Hardenbergstrasse 36, Berlin 10623, Germany}
\end{center}

\vspace{0.5cm}

\twocolumngrid

\section{Brief review of the theory of Bose selection}

For the manuscript to be self-contained, we here briefly recapitulate the derivation of the mean-field rate Eqs. and of the Bose-selection (BS) conditions [Eq.~(2)] used in the main text, which were derived in Refs.~\cite{Vorberg2013, Vorberg2015}. Following Eq.~(1), a dissipative Bose gas coupled weakly to a generic bath is described, in the interaction picture, by the Lindblad master equation
\begin{equation} \label{eq:lindgen}
\frac{d\hat{\rho}(t)}{dt} = \sum_{i,j} R_{ij}\left[\hat{L}_{ij} \hat{\rho}(t)\hat{L}_{ij}^\dagger - \frac{1}{2}\left\{\hat{L}_{ij}^\dagger \hat{L}_{ij}, \hat{\rho}(t) \right\}  \right],
\end{equation}
where the operators $\hat{L}_{ij}=\hat{c}_i^\dagger \hat{c}_j$ describe a quantum jump from the single-particle eigenstate $\ket{j}=\hat{c}_j^\dagger\ket{v}$ to $\ket{i}=\hat{c}_i^\dagger \ket{v}$,  occurring at rate $R_{ij}$. Here, $\ket{v}$ represents the vacuum state. For ease of notation, in this section we drop the subscript $S$ for the system density matrix $\hat{\rho}$, as compared to Eq.~(1). We denote with $\hat{n}_i=\hat{c}_i^\dagger \hat{c}_i$ the number operators. Equation~\eqref{eq:lindgen} generically describes the dynamics of the system coupled to an arbitrary environment, being it thermal or engineered, in the ultraweak-coupling limit where the full secular (rotating-wave) approximation is valid~\cite{Breuer2007}, namely such that the system-environment coupling is much smaller than the gaps in the system. Eq.~\eqref{eq:lindgen} implies a decoupling of the equations for the populations $p_{\bm{n}}=\bra{\bm{n}}\hat{\rho}(t)\ket{\bm{n}}$ in the Fock basis $\ket{\bm{n}}$ of eigenstate occupations and for their coherences. Since the coherences fully decay, the steady state is diagonal in the Fock basis, $\hat{\rho}_{ss}=\sum_{\bm{n}}p_{\bm{n}}\ket{\bm{n}}\!\bra{\bm{n}}$, and thus only the population evolution needs be solved to determine it. The latter follows from~\eqref{eq:lindgen} as~\cite{Vorberg2015}
\begin{equation}
\label{eq:populations}
\frac{d p_{\bm{n}}(t)}{dt} = \sum_{i,j} n_i(n_j + 1)\big[R_{ij}p_{\bm{n}_{ji}}(t) - R_{ji}p_{\bm{n}}(t)\big]\ ,
\end{equation}
where $\bm{n}_{ji} = (n_1, \ldots, n_{i}-1, \ldots, n_{j}+1,\ldots)$, and it features the bosonically enhanced many-body rates $R_{ij}n_i(n_j+1)$ and $R_{ji}n_i(n_j+1)$. Similarly, one can derive the evolution equation $d n_i(t)/dt = \mathrm{tr}\left[\hat{n}_i(t) d \hat{\rho}(t)/dt\right]$ for the mean many-body occupations $n_i(t)=\mathrm{tr}[\hat{\rho}(t)\hat{n}_i]$, obtaining
\begin{multline}
\frac{d n_i(t)}{dt} = \sum_{j} \Big\{R_{ij} [n_j(t) + \langle \hat{n}_i\hat{n}_j\rangle (t)] \\
 - R_{ji} [n_i(t) + \langle \hat{n}_i\hat{n}_j\rangle (t)]\Big\}\ ,
 \label{eq:mfwcorr}
\end{multline}
where we indicated with $\langle \hat{n}_i\hat{n}_j\rangle$ the two-particle correlation functions $\langle \hat{n}_i\hat{n}_j\rangle=\mathrm{tr}[\hat{n}_i\hat{n}_j \hat{\rho}]$. Eq.~\eqref{eq:mfwcorr} represents the first step of an infinite hierarchy of Eqs.~for the correlation functions, where the Eq.~for the $i$-th order correlation function depends on the $(i+1)$th order correlation functions, and cannot thus be solved exactly. The hierarchy can be truncated by means of a mean-field approximation, consisting in assuming that the two-particle correlation functions factorize, thus neglecting two-particle correlations, that is
\begin{equation} 
\langle \hat{n}_i\hat{n}_j\rangle \simeq n_i n_j.
\end{equation}
This, in turn, corresponds to assuming a Gaussian ansatz for the system's state, of the form $\hat{\rho}\propto \exp(-\sum_{i}\eta_{i}\hat{n}_i^\dagger)$ for which Wick's decomposition applies~\cite{Vorberg2015}. Within this approximation, one obtains the mean-field Eqs.
\begin{multline}
\frac{d n_i(t)}{dt} = \sum_{j} \Big\{ R_{ij}n_j(t) [1 + {n}_i(t)] \\
 - R_{ji} n_i(t) [ 1+{n}_j(t)]\Big\}\ . 
 \label{eq:meanfield1}
\end{multline}
To study the steady state in the Bose-selection regime involving large particle numbers $N$, we consider the asymptotic limit $dn_i/dt=0$ and a formal high-density expansion of the mean occupations,
\begin{equation}
n_i = \nu_i N + \sum_{j=1}^{+\infty} \nu_i^{(j)} N^{j-1}\ .
\end{equation}
Inserting into Eq.~\eqref{eq:meanfield1}, and requiring that in leading order $n_i>0$, one finds that the occupations of the selected states (forming a set $S$) are ruled by 
\begin{equation}
\label{eq:condition1}
\sum_{j\in S} A_{ij} \nu_j = 0, \qquad \nu_j>0,\qquad \qquad( i \in S)
\end{equation}
involving the rate-asymmetry matrix $A_{ij}= R_{ij}- R_{ji}$. The occupation of the non-selected states (forming a set $\bar{S}$) is determined primarily by the occupation of selected states (similarly to a Bogoliubov approximation) and is approximated by  
\begin{equation} \label{eq:depletedpop}
\nu^{(1)}_i\approx  -\sum_{j\in S} R_{ij} \nu_j/\sum_{j \in S}A_{ij} \nu_j >0.
\end{equation}
The physical condition that these occupations are non-negative implies
\begin{equation} \label{eq:condition2}
\sum_{j \in S}A_{ij} \nu_j < 0,\qquad( i \in \bar{S}).
\end{equation}
Together, Eqs.~\eqref{eq:condition1} and Eq.~\eqref{eq:condition2} give the Bose-selection conditions of Eq.~(2) of the main text.
It has been proven that these always single out a unique set $S$ of selected states~\cite{Vorberg2013, Vorberg2015}.

\section{Structure of the asymmetry matrix}

\begin{figure}
\centering
\includegraphics[width=\linewidth]{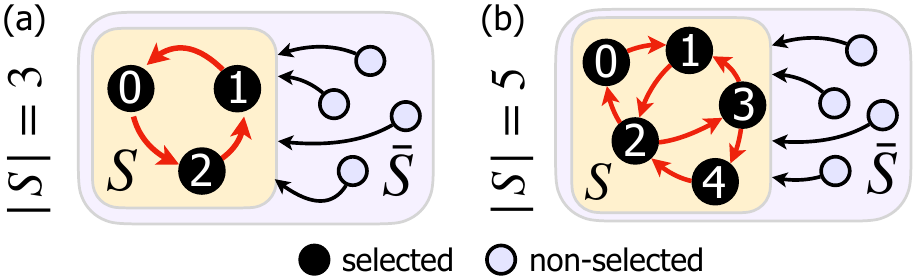}
\caption{Asymmetry-matrix structures compatible with Bose selection into three states (a) and five states (b).}
\label{fig:connectivity}
\end{figure}

In this Appendix, we show that the connectivities of Fig.~1(b) [reported in Fig.~\ref{fig:connectivity}] satisfy the asymptotic condensation conditions of Eq.~(2), thus providing some intuition about the structures of the matrix $A$ admitting Bose selection. These examples were constructed \textit{ad hoc} to provide a crisp illustration of specific features possessed by possible solutions of the Bose-selection conditions. Because of this, they are also an over-simplified description of realistic networks, which usually involve all-to-all couplings. 

The first property of these examples is that all non-selected states directly feed the set $S$ of selected states: this is chosen, such that the condensation condition $A_{\bar{S}S}\bm{\nu}_S<0$ is trivially fulfilled: all matrix elements $A_{\bar{S}S}$ are negative, while $\bm{\nu}_S>0$. In this case, it is also unnecessary to specify which non-selected state feeds which selected state: the above inequalities are solved in all cases. Differently from the ideal networks constructed here, in a realistic network also outflow from $S$ to $\bar{S}$ is present, making the solution of the inequalities highly non-trivial and requiring numerical methods. Another simplification comes from the fact that non-selected states will also be connected among themselves. This coupling can however be neglected in leading order in the high-density expansion, since the selection pattern and the residual occupation of non-selected states is primarily determined by their coupling to the selected states~\cite{Vorberg2015}, as also shown in Eq.~\eqref{eq:depletedpop}. More generally, if a network does not feature all-to-all couplings or if the asymmetry matrix $A$ contains zeros, care must be taken because (i) multiple steady states may exist and (ii) even for a single steady state, the actual steady state may be fully determined only by considering also higher-order terms in the high-density expansion, neglected in the derivation of the BS conditions based on $A$. In all cases considered in the manuscript, the asymmetry matrices feature non-zero $A_{ij}$, such that these problems are not encountered.

Assuming from now on that the condition $A_{\bar{S}S}n_S<0$ is fulfilled, we proceed in showing that the connectivities within $S$ in Fig.~\ref{fig:connectivity} sustain Bose selection. A non-trivial solution $\bm{\nu}_S$ of the Bose-selection condition $A_{SS}\bm{\nu}_S=0$ with $\nu_{S}>0$, up to normalization, has the general form~\cite{Knebel2013,Knebel2015}
\begin{equation} \label{eq:n_pf}
\bm{\nu}_S = \left[\mathrm{Pf}(A_{SS}^{[0]}),-\mathrm{Pf}(A_{SS}^{[1]}), \dots , \mathrm{Pf}(A_{SS}^{[|S|-1]})\right],
\end{equation}
where $A_{SS}^{[j]}$ is the matrix obtained from $A_{SS}$ by removing the $j$th row and column, and where $\mathrm{Pf}(A)$ is the Pfaffian,
\begin{align}
\mathrm{Pf}(A) & = \frac{1}{2^n n!}\sum_{i_1,j_1, \dots}\varepsilon_{i_1j_1i_2j_2\dots}A_{i_1 j_1} A_{i_2 j_2}\dots A_{i_M, j_M}, 
\end{align}
which can also be computed via
\begin{align} \label{eq:pfaff}
\mathrm{Pf}(A)& = \sum_{j=2}^{N} (-1)^j a_{1,j} \mathrm{Pf}(A_{S}^{[1,j]}).
\end{align}
The notation $A_{SS}^{[i,j]}$ indicates the matrix obtained from $A_{SS}$ by removing the $i$th and $j$th rows and columns.
Therefore, selection in $S$ is possible if the submatrices $A_{SS}^{[j]}$ have negative (positive) Pfaffian when $j$ is odd (even). In general, whether this occurs depends not only on the network connectivity and topology (e.g., whether loops are present), but also on the specific values of the rate-asymmetries $A_{ij}$~\cite{Geiger2018}. The occupation fractions $\nu_{S,i}= |\mathrm{Pf}(A_{SS}^{[i]})|$ depend non-linearly on the entries of $A_{SS}$. \\

 {\bf Three-state selection.} For selection into three states, the conditions $A_{SS}=0$ and $\bm{\nu}_S>0$ from Eq.~(2) explicitly read
\begin{equation}
\begin{pmatrix}
0 & A_{01} & A_{02}\\
-A_{01} & 0 & A_{12}\\
-A_{02} & -A_{12} & 0
\end{pmatrix} \begin{pmatrix}
\nu_{S,0} \\
\nu_{S,1} \\
\nu_{S,2}
\end{pmatrix} = 0\ , \qquad \nu_{S, i} > 0.
\end{equation}
One then finds that the system of equalities admits only the solution
\begin{equation}
A_{01} = {\nu}_{S,2},\qquad A_{02} = -{\nu}_{S,1}, \qquad A_{12} = {\nu}_{S,0} \ ,
\end{equation}
up to normalization. The matrix $A_{SS}$ must thus have the loop geometry depicted in Fig.~1(b)-top and reported with labelling in Fig.~\ref{fig:connectivity}(a). For three-state selection only, the matrix elements of $A_{SS}$ directly also give the relative occupations in the condensate vector.  \\

{\bf Five-state selection states.} 
While for three-state selection the loop geometry of the rate asymmetries is the only one consistent with the condensation conditions, the situation becomes more complex for selection into larger sets of states. The five-sites connectivity of Fig.~1(b)-bottom (reported with labelling in Fig.~\ref{fig:connectivity}), realized by the protocol of Fig.~3(a), gives rise to the following asymmetry within $S$
\begin{equation}
A_{SS} = \begin{pmatrix}
0 & -A_{01} & A_{02}&0& 0\\
A_{01} & 0 & -A_{12} & A_{13} & 0 \\
-A_{02}& A_{12} & 0 & -A_{23}& A_{24}  \\
0& -A_{13}& A_{23} & 0 & -A_{34}  \\
0&  0& -A_{24}& A_{34}& 0 
\end{pmatrix}, \label{eq:Aform}
\end{equation}
with positive $A_{i,j}>0$. 
From Eq.~\eqref{eq:n_pf} and Eq.~\eqref{eq:pfaff}, the condensation vector is then related to the matrix elements of $A$ by
\begin{align} \label{eq:nS5}
\bm{\nu}_S =&  [A_{12}A_{34} - A_{13}A_{24}, \quad A_{02}A_{34}, \quad A_{01}A_{34},\nonumber \\
& A_{01}A_{24}, \quad A_{01}A_{23}-A_{02} A_{13}].
\end{align}
Hence, positive occupation is guaranteed provided the constraints
\begin{equation} 
\label{eq:fiveconn}
A_{12} A_{34} > A_{13} A_{24}, \qquad A_{01} A_{23}> A_{02} A_{13},
\end{equation}
are fulfilled. This is indeed the case, for our choice of the sign $A_{ij}>0$, while, for generic signs, one would need to explicitly solve Eq.~\eqref{eq:fiveconn}. 

\section{Numerical simulation methods} 
\label{sec:nummeth}

\begin{figure}
\centering
\includegraphics[width=\linewidth]{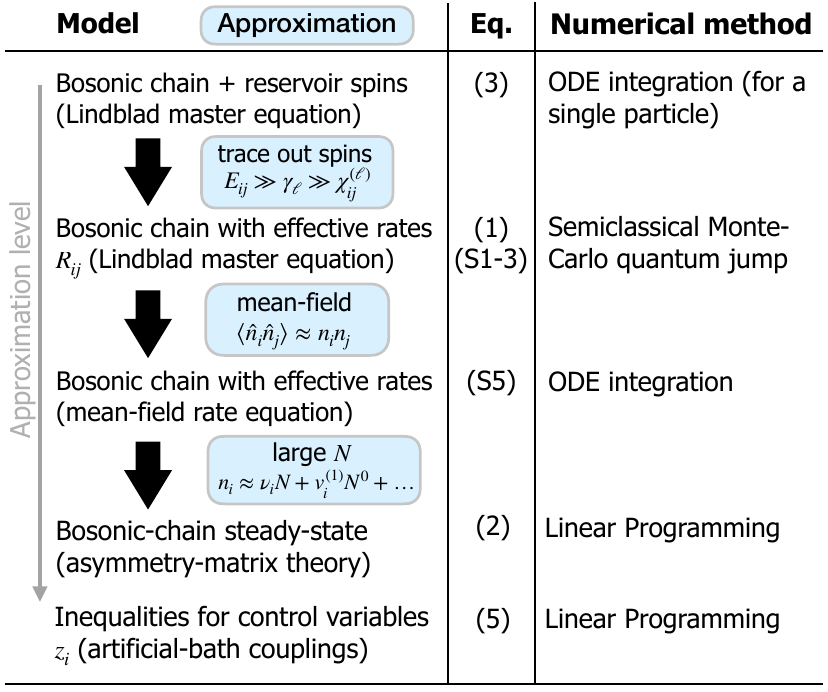}
\caption{Diagram of the central models, Eqs., and corresponding numerical method used to solve them.}
\label{fig:diagram}
\end{figure}

We here discuss the different numerical methods used in simulations. Fig.~\ref{fig:diagram} summarizes the central models, the approximations they entail, and the numerical method used to solve them.\\

 \textbf{Many-body master equation}. To simulate the evolution of the mean occupations from the master Eq.~(1) [also reported in~\eqref{eq:lindgen}], we adopt the semiclassical Monte-Carlo quantum-jump method of Refs.~\cite{Vorberg2013, Vorberg2015}. Equation~\eqref{eq:populations} for the many-body probabilities, descending from \eqref{eq:lindgen}, can be unraveled in terms of a $N$-particle random walk in Fock space. The mean occupations are thus computed by averaging over a large number of stochastic trajectories simulated through Gillespie's algorithm, as detailed in Ref.~\cite{Vorberg2015}.\\ 
 
\textbf{Mean-field equations} The mean-field rate equations, ~\eqref{eq:meanfield1}, are solved by time integration with an ODE solver. The steady state is found via long-time propagation while monitoring the norm of $d\bm{n}/dt$.\\

\textbf{Large-N asymptotic theory}. Solutions of the Bose-selection inequalities~(2), which predict the steady state for large particle numbers, are solved via the linear programming method of Ref.~\cite{Knebel2015}. To this end, we show that the inequalities (2), which we restate here,
\begin{equation} \label{eq:age_cond}
A_{SS}\bm{\nu}_S = 0, \quad A_{\bar{S}S} \bm{\nu}_S < 0, \quad \bm{\nu}_S>0, \quad \bm{\nu}_{\bar{S}}=0 ,
\end{equation}
 can be recast in the form used in Refs.~\cite{Knebel2015, Tucker1957}, which reads
\begin{equation} \label{eq:condknebel}
A \bm{\nu} \le 0, \qquad (1 - A) \bm{\nu} > 0, \qquad \bm{\nu} \ge 0.
\end{equation} 
Consider a solution of Eq.~\eqref{eq:condknebel}. Since $\bm{\nu}\ge0$, the entries of $\bm{\nu}$ can be divided into two sets, namely the set of non-zero entries $S=\{i : \nu_i>0\}$ and the set of vanishing entries $\bar{S}=\{i: \nu_i=0\}$. Identifying $S$ with the set of selected states, this gives the third and fourth condition in Eq.~\eqref{eq:age_cond}. Accordingly, we decompose $\bm{\nu}=[\bm{\nu}_S, \bm{\nu}_{\bar{S}}] = [\bm{\nu}_S,\bm{0}]$ and the matrix $A$ gets decomposed into matrix blocks $\{A_{SS}, A_{S\bar{S}}, A_{\bar{S}S}, A_{\bar{S}\bar{S}}\}$. Then, the system of inequalities~\eqref{eq:condknebel} gives the following conditions for the matrix blocks and $\bm{\nu}_S$,
\begin{eqnarray} 
 A_{SS}\bm{\nu}_S & \le 0, \nonumber \\
 A_{\bar{S}S}\bm{\nu}_S & <0, \nonumber \\
 \bm{\nu}_S-A_{SS} \bm{\nu}_S & > 0, \nonumber \\
 \bm{\nu}_S & >0 .
\label{eq:halfway}
\end{eqnarray}
The second inequality in~\eqref{eq:halfway} is the same as the second inequality in \eqref{eq:age_cond}. We next note that, since $A_{SS}$ is skew-symmetric, its diagonal entries vanish and it thus follows that $\bm{\nu}_S^T A_{SS} \bm{\nu}_S = 0$. However, the multiplication by the strictly positive $\bm{\nu}_S^T$ from the left-hand side cannot have changed the nature of the latter equality (e.g., turn an inequality into an equality), and it must hold that $A_{SS} \bm{\nu}_S = 0$. The second condition in Eq.~\eqref{eq:age_cond} is then also recovered. This shows that the solution of Eq.~\eqref{eq:condknebel} fulfils the BS conditions~\eqref{eq:age_cond}. The converse is obtained straightforwardly by noting that the solution of~\eqref{eq:age_cond} directly fulfils Eq.~\eqref{eq:halfway}. \\

\section{Inequalities for the control variables}

To turn the BS conditions into a control problem in terms of a number of control variables, leading to Eq.~(7), we consider a decomposition of the total asymmetry matrix $A$ in the form $A=\sum_\ell z_\ell A^{(\ell)}$, as given by a controllable coupling to a number of baths, where $\bm{z}$ is the vector of control variables. Inserting this decomposition and a target condensation vector $\bm{\nu}_S$ into the BS conditions gives
\begin{subequations} \label{eq:interstep1}
\begin{align}
 \sum_{j \in S} \sum_\ell z_\ell A_{ij}^{(\ell)}\nu_j & = 0,\qquad i\in S \\
 \sum_{j\in S}\sum_\ell z_\ell A_{ij}^{(\ell)} \nu_j & < 0,\qquad i \in \bar{S},
\end{align}
\end{subequations}
where the conditions $\bm{\nu}_S>0$ and $\bm{\nu}_{\bar S}=0$ have been imposed by construction through the choice of the target $\bm{\nu}$. We interpret the conditions in Eq.~\eqref{eq:interstep1} as constraints on $\bm{z}$, yielding the inequalities of Eq.~(7), namely
\begin{equation}
(B\bm{z})_{\bar S} < 0, \qquad (B\bm{z})_S = 0,
\end{equation}
where we defined the matrix $B$ given by
\begin{equation} \label{eq:condcontr1}
B_{i\ell} = \sum_{j} A_{ij}^{(\ell)}\nu_j\ .
\end{equation}
Similarly to the BS conditions~(2), the existence of solutions $\bm{z}$ for the control inequalities of Eq.~\eqref{eq:condcontr1} can be efficiently verified via linear programming. To this end, it is convenient to reformulate Eq.~\ref{eq:condcontr1} using the modified BS conditions of Eq.~\eqref{eq:condknebel}, which, we proved in Sec.~\ref{sec:nummeth}, are equivalent to Eq.~(2). Again inserting the decomposition for $A$ and a target vector $\nu_S$, one finds
\begin{equation} \label{eq:Blin}
B \bm{z} \le 0, \quad B \bm{z} < \bm{\nu}.
\end{equation}
These inequalities are equivalent to Eq.~(7), but more amenable for numerical solution. Indeed, we can find solutions of the system~\eqref{eq:Blin} by representing it as the constraint in the linear program
\begin{align} \label{eq:lp}
\min_{\bm{z}} f(\bm{z}) \quad \text{ subject to }\quad &\mathcal{B} \bm{z} \le \bm{n}_\epsilon, \nonumber \\ & -z_{\rm max} \le z_j \le z_\mathrm{max},
\end{align}
where $f(\bm{z})$ is a linear function of $\bm{z}$. While the goal of the linear program is to optimize the linear function $f(\bm{z})$ compatibly with the constraints, for our purposes it is sufficient to find a feasible solution satisfying the latter, and we can thus simply choose $f(\bm{z})=0$. In~\eqref{eq:lp}, we have defined the matrix $\mathcal{B}$ with block structure $\mathcal{B} = [B, B]$ and the vector $\bm{n}_\epsilon = [\bm{0}_{\mathrm{dim}(B)}, \bm{\nu} - \bm{\epsilon}]$, where $\bm{\epsilon}$ are tolerance parameters with chosen value $\epsilon_j=10^{-2}$, allowing us to include the strict inequality of the second condition in~\eqref{eq:Blin} as a `$\le$' inequality in~\eqref{eq:lp}. The constraint on $\bm{z}$ is imposed with $z_{\mathrm{max}} = 1$ to ensure experimental practicality of the parameters found, such that the dispersive coupling does not exceed a maximum value $\chi_{\rm max}$, as discussed in the main text. The linear program of Eq.~\eqref{eq:lp} can then be solved by means of SciPy built-in routines~\cite{Scipy}.

\section{Reservoir engineering with driven-damped spins}
\label{app:spinreservoirs}

In this section, further details about the derivation of the rates $R_{ij}=\sum_\ell R_{ij}^{(\ell)}$ produced by the driven-damped spins used as artificial reservoirs are discussed. Following the main text, we highlight the approximations involved. Considering a single spin reservoir for simplicity, the Hamiltonian, in a frame rotating at the driving frequency of the spin, reads
\begin{equation}
\hat{H} = \hat{H}_S + \chi_\ell \hat{a}_\ell^\dagger \hat{a}_\ell \hat{\sigma}_\ell^z + \frac{\delta_\ell}{2} [\hat{\sigma}_\ell^z + r_\ell \hat{\sigma}_\ell^x]\ ,
\end{equation}
where $\delta_\ell$ is the spin-drive detuning, $r_\ell$ is the ratio between the Rabi frequency of the drive and its detuning, and $\hat{H}_S$ is the bosonic-chain Hamiltonian,
\begin{equation}
\hat{H}_S = \sum_{\ell=0}^{M-1} \omega_\ell \hat{a}_\ell^\dagger \hat{a}_\ell - J\sum_{\ell=0}^{M-2} (\hat{a}_{\ell} \hat{a}_{\ell+1}^\dagger + \hat{a}_{\ell}^\dagger \hat{a}_{\ell+1} )\ .
\end{equation}
The dissipative dynamics of the system is described by the master equation
\begin{equation} \label{eq:resspinme}
\frac{d\hat{\rho}}{dt} = -i[\hat{H}, \hat{\rho}] + \gamma_\ell D[\hat{\sigma}_\ell^-]\hat{\rho}\ ,
\end{equation}
for the density matrix $\hat{\rho}$ of the coupled resonators-spins system, where $\gamma_\ell$ is the relaxation rate of the spin and $D[\cdot]\hat{\rho}$ is a Lindblad dissipator, $D[\cdot]\hat{\rho} = (\cdot)\hat{\rho}(\cdot)^\dagger - (1/2)\{(\cdot)^\dagger(\cdot), \, \hat{\rho}\}$. By diagonalizing the spin Hamiltonian via the unitary transformation $\hat{U}_\ell=-\sin(\theta_\ell)\hat{\sigma}_\ell^z + \cos(\theta_\ell)\hat{\sigma}_\ell^x$ with $\theta_\ell = \arctan(r_\ell)/2$, one obtains
\begin{equation}
\hat{H} = \hat{H}_S -\frac{\delta_\ell\chi_\ell}{\mathcal{E}_\ell} \hat{a}_\ell^\dagger \hat{a}_\ell \big[\hat{\sigma}_\ell^z + r \hat{\sigma}_\ell^x\big] -  \frac{\mathcal{E}_\ell}{2} \hat{\sigma}_\ell^z\ ,
\end{equation}
with $\mathcal{E}_\ell =\delta_\ell \sqrt{1 + r_\ell^2}$. The transformation also affects the dissipator. It transforms as 
\begin{equation}
\gamma_\ell D[\hat{U}_\ell^\dagger \hat{\sigma}_\ell^- \hat{U}_\ell]\hat{\rho} = \gamma_\ell \! \! \! \! \sum_{\substack{\alpha, \beta\\ \in \{z, +, -\}}}\! \! \! \! K^{(\ell)}_{\alpha\beta} \big(\hat{\sigma}_\ell^\alpha \hat{\rho} [\hat{\sigma}_\ell^\beta]^\dagger -\frac{1}{2}\{[\hat{\sigma}_\ell^\beta]^\dagger \hat{\sigma}_\ell^\alpha , \hat{\rho}\} \big)\ ,
\end{equation}
where the Kossakowski matrix $K^{(\ell)}$ reads
\begin{equation}
K^{(\ell)} = \begin{bmatrix}
\frac{r_\ell^2}{4(1+r_\ell^2)} & -\frac{r_\ell \cos^2\theta_\ell}{2(1+r_\ell^2)} & \frac{r_\ell \sin^2\theta_\ell}{2(1+r_\ell^2)} \\
-\frac{r_\ell \cos^2\theta_\ell}{2(1+r_\ell^2)} & \cos^4 \theta_\ell & \frac{r_\ell^2}{4(1+r_\ell^2)} \\
\frac{r_\ell \sin^2\theta_\ell}{2(1+r_\ell^2)} & - \frac{r_\ell^2}{4(1+r_\ell^2)} & \sin^4 \theta_\ell
\end{bmatrix}\ .
\end{equation}
In the limit where the level splitting $|\mathcal{E}_\ell|$ of the spin is much larger than the decay rate $\gamma_\ell$, that is 
\begin{equation}
|\mathcal{E}_\ell| \gg \gamma_\ell\ ,
\end{equation}
terms corresponding to off-diagonal elements of the Kossakowski matrix are off-resonant and can be neglected in rotating-wave approximation (RWA) (\textbf{Approximation I}). The coupling of the spins to the system, as detailed below, will be chosen to be much weaker than $\gamma_\ell$, such that it does not challenge the validity of this approximation. The transformed dissipator then reduces to  
\begin{equation}
 \gamma_\ell^+ D[\hat{\sigma}_\ell^+]\hat{\rho} + \gamma_\ell^- D[\hat{\sigma}_\ell^-]\hat{\rho} + \gamma_\ell^z D[\hat{\sigma}_\ell^z]\hat{\rho}\ ,
\end{equation}
with dressed decay rates
\begin{align}
 & \gamma_\ell^+ = \gamma_\ell \cos^4(\theta_\ell)\ ,\\
 & \gamma_\ell^- = \gamma_\ell \sin^4(\theta_\ell)\ , \\
 & \gamma_\ell^z = \frac{\gamma_\ell}{4}\frac{r_\ell^2}{1 + r_\ell^2}\ .
\end{align}
as given in Eq.~(4) of the main text.
When the spin level splitting $\mathcal{E}_\ell$ is in resonance with a level spacing in the system, and the spin decay rate is larger than the effective system-spin coupling, the spin can be traced out, yielding effective transition rates between the bosonic-chain eigenstates, ruled by a Lorentzian bath spectral density. To show this, we start by formally diagonalizing the bosonic chain, $\hat{H}_S=\sum_i E_i \hat{c}_i^\dagger \hat{c}_{i}$ with $\ket{i}=\hat{c}_i^\dagger \ket{v}$ the single-particle eigenstates and $\ket{v}$ the vacuum. In the interaction picture with respect to the bare energies of the bosonic chain and spin, the Hamiltonian reads
\begin{multline} \label{eq:anontherH}
\hat{H}(t) = -\sum_{i>j} \big( \chi^{(\ell)}_{ij}e^{i E_{ij} t}\hat{c}^\dagger_i \hat{c}_{j} + \chi^{(\ell)*}_{ij}e^{-i E_{ij} t}\hat{c}^\dagger_j \hat{c}_{i}  \big)\\
\times \big(\frac{1}{r_\ell}\hat{\sigma}_\ell^z +  e^{-i \mathcal{E}_\ell t} \hat{\sigma}_\ell^+
 +  e^{i \mathcal{E}_\ell t}\hat{\sigma}_\ell^- \big) ,
\end{multline}
where $E_{ij}=E_i-E_j$, and 
\begin{equation}
\chi^{(\ell)}_{ij}=  \frac{\chi_\ell r_\ell M_{ij}^{(\ell)}}{\sqrt{1+r_\ell^2}} \ ,
\end{equation}
with matrix element $M_{ij}^{(\ell)} = \bra{i} \hat{a}_\ell^\dagger \hat{a}_\ell\ket{j}$. The dissipator remains unchanged. 
Consider now the case in which $\mathcal{E}_\ell\approx E_{ij}>0$, namely the spin level splitting is (quasi)resonant with the level spacing $E_{ij}$. The terms $\hat{c}_i^\dagger \hat{c}_{j}\hat{\sigma}_\ell^+$ and $\hat{c}_{j}^\dagger \hat{c}_i\hat{\sigma}_\ell^-$ become resonant, and other terms can be neglected, in RWA, if the spacing is much larger than the effective couplings (\textbf{Approximation II}),
\begin{equation}
E_{ij} \gg |\chi_{ij}^{(\ell)}| .
\end{equation}
This yields the Schr\"odinger's picture Hamiltonian (up to a constant energy shift)
\begin{align}
\hat{H} = & \frac{E_{ij}}{2} (\hat{c}_i^\dagger \hat{c}_i - \hat{c}_j^\dagger\hat{c}_j) - \frac{\mathcal{E}_\ell}{2}\hat{\sigma}_\ell^z \nonumber \\
& -\sum_{(i,j)\mathrm{ res} }  \big( \chi^{(\ell)}_{ij}\hat{c}^\dagger_i \hat{c}_{j} \hat{\sigma}_\ell^+ + \chi^{(\ell)*}_{ij}\hat{c}^\dagger_j \hat{c}_{i}\hat{\sigma}_\ell^-  \big) .
 \label{eq:anotherRWAH}
\end{align}
The summation runs over all pairs of single-particle eigenstates with energy spacing quasiresonant with $\mathcal{E}_\ell$. For ease of notation, in the following we focus on a single such pair.

Next, we will trace out the spin, treating it as an environment degree of freedom, such that $\hat{H}$ of Eq.~\eqref{eq:anotherRWAH} represents a system-bath coupling Hamiltonian. We proceed following standard Born-Markov-secular derivations~\cite{Breuer2007}, as discussed in the main text. The dynamics of the bosonic system is described by the Markovian quantum master equation for its density matrix $\hat{\rho}_S$~\cite{Breuer2007},
\begin{equation} \label{eq:markovianme}
\frac{d\hat{\rho}_S}{dt} = \int_0^{+\infty} ds \ \mathrm{tr}_\sigma\big[ \hat{H}(t), [\hat{H}(t-s), \hat{\rho}_S\otimes \hat{\rho}_\sigma]\big],
\end{equation}
where $\mathrm{tr}_\sigma$ and $\hat{\rho}_\sigma$ denote the trace over the degrees of freedom of the spin and the spin's steady-state density matrix, respectively. $\hat{H}(t)$ represents the Hamiltonian of Eq.~\eqref{eq:anotherRWAH} in the interaction picture. By adopting Eq.~\eqref{eq:markovianme}, we are assuming the Born and Markov approximations (\textbf{Approximation III}). The latter are justified, if the spin state is negligibly affected by the coupling to the system, such that the states of the system (resonators) and the bath (spins) can be approximated as factorized, $|\chi_{ij}^{(\ell)}| \ll |\mathcal{E}_\ell|.$
Moreover, the spin must relax quickly enough to its steady state, such that its state can be considered as constant over the evolution timescales of the system. These conditions imply that the decay rate of the spin, $\gamma_\ell$, needs to be much stronger than the effective spin-resonator coupling entering the Hamiltonian~\eqref{eq:anotherRWAH},
\begin{equation}
|\chi_{ij}^{(\ell)}| \ll \gamma_\ell \ .
\end{equation}
Note, however, that $\gamma_\ell$ needs also be small enough for Approximations I and II to be valid, and thus a hierarchical separation of energy scales between the level splittings, $\gamma_\ell$ and the effective system-spin coupling must be fulfilled.

Inserting Eq.~\eqref{eq:anotherRWAH} into~\eqref{eq:markovianme} and following standard manipulations~\cite{Breuer2007}, a master equation in the Lindblad form as in Eq.~(1) is obtained,
\begin{equation} \label{eq:meS}
\frac{d\hat{\rho}_S}{dt}  = R_{ij}^{(\ell)} D[\hat{L}_{ij}]\hat{\rho}_S+  R_{ji}^{(\ell)} D[\hat{L}^\dagger_{ij}]\hat{\rho}_S \ ,
\end{equation}
where Lamb-shift corrections have been neglected, and the rates $R_{ij}^{(\ell)}$ read
\begin{align} \label{eq:rates_posdel}
&  R^{(\ell)}_{ij}= |\chi_{ij}^{(\ell)}|^2 \mathcal{S}_{\ell}^+(-E_{ij}),\nonumber \\
& R^{(\ell)}_{ji}= |\chi_{ij}^{(\ell)}|^2 \mathcal{S}_\ell^-(E_{ij}),  \quad (\mbox{for }  \delta_\ell>0)\ . 
\end{align}
The quantum noise spectra $\mathcal{S}_\ell^{\pm}(\omega)$ in Eq.~\eqref{eq:rates_posdel} are defined by
\begin{align}
\mathcal{S}_\ell^\pm(\omega) & = \int_{-\infty}^{\infty}dt e^{i\omega t} \langle [\hat{\sigma}_\ell^\pm(t)]^\dagger \hat{\sigma}_\ell^\pm(0)\rangle_{\sigma}, \nonumber \\
&  = 2\mathrm{Re}\int_{0}^{\infty}dt e^{i\omega t} \langle [\hat{\sigma}_\ell^\pm(t)]^\dagger \hat{\sigma}_\ell^\pm(0)\rangle_{\sigma} . \label{eq:spectrum}
\end{align}
Here, $\langle \cdot \rangle_{\sigma} = \mathrm{tr}[(\cdot) \rho_{\sigma}]$ indicates the average with respect to the spin's steady state $\hat{\rho}_{\sigma}$. For negative times we have used the property $\langle[\sigma_\ell^\pm(-t)]^\dagger \sigma_\ell^\pm(0) \rangle_{\sigma} = \langle[\sigma_\ell^\pm(t)]^\dagger \sigma_\ell^\pm(0) \rangle_{\sigma}^*$.
The time-dependent operators $\hat{\sigma}_\ell^+(t)$ are interaction-picture representations of $\hat{\sigma}_\ell^\pm$. In deriving Eq.~\eqref{eq:meS}, an additional secular approximation has been made to arrive at the Lindblad form, which is consistent with Approximation II. The master Eq.~\eqref{eq:meS} has been derived by assuming quasiresonance between $\mathcal{E}_\ell$ and $E_{ij}>0$, implying a positive detuning $\delta_\ell > 0$. For negative detuning, the terms $\hat{c}_i^\dagger \hat{c}_{j}\hat{\sigma}_\ell^-$ and $\hat{c}_{j}^\dagger \hat{c}_i\hat{\sigma}_\ell^+$ in Eq.~\eqref{eq:anontherH} become resonant instead. In this case, the derivation of the master equation is the same as above, only with the role of $\hat{\sigma}_\ell^-$ and $\hat{\sigma}_\ell^+$ [and thus $\mathcal{S}_\ell^-(\omega)$ and $\mathcal{S}_\ell^+(\omega)$] exchanged, yielding rates
\begin{align}
 & R^{(\ell)}_{ij}= | \chi_{ij}^{(\ell)}|^2 \mathcal{S}_\ell^-(-E_{ij}), \\
& R^{(\ell)}_{ji}= | \chi_{ij}^{(\ell)}|^2 \mathcal{S}_\ell^+(E_{ij}),   \quad (\mbox{for } \delta_\ell < 0) .
 \end{align}
 
 To explicitly compute the spectra~\eqref{eq:spectrum} entering the rates, the spin's steady state and correlation functions are needed. Consistently with the Markov approximation, we treat the spin as negligibly influenced by the coupling to the system, and compute such quantities as they are determined by the free decay and the coherent drive. This is achieved by solving the optical Bloch equations and using the quantum regression theorem~\cite{Carmichael1999, Breuer2007}. The problem is formally equivalent to that of a spontaneously decaying qubit with additional dephasing. The optical Bloch equations read
\begin{equation} \label{eq:Bloch}
\frac{d}{dt} \begin{bmatrix}
\langle \sigma_\ell^z\rangle(t) \\
\langle \sigma_\ell^+\rangle(t) \\
\langle \sigma_\ell^-\rangle(t) \\
\end{bmatrix} = 
\begin{bmatrix}
-(\gamma_\ell^+ + \gamma_\ell^-) \langle \sigma_\ell^z\rangle(t) + \gamma_\ell^+ - \gamma_\ell^- \\
(-i\mathcal{E}_\ell-\Gamma_\ell)\langle \sigma_\ell^+\rangle(t) \\
(i\mathcal{E}_\ell-\Gamma_\ell) \langle \sigma_\ell^-\rangle(t)
\end{bmatrix} ,
\end{equation}
where 
\begin{equation}
\label{eq:Gammaell}
\Gamma_\ell = \frac{\gamma_\ell^+ + \gamma_\ell^-}{2} + 2 \gamma_\ell^z = \frac{\gamma_\ell}{4}\left[3 - \frac{1}{1+r_\ell^2}\right] .
\end{equation}
 The steady state $\hat{\rho}_\sigma$ thus satisfies
\begin{align}
\langle \hat{\sigma}_\ell^+ \rangle_{\sigma} = \langle \hat{\sigma}_\ell^-\rangle_{\sigma} = 0,\qquad 
\langle \hat{\sigma}_\ell^z \rangle_{\sigma} = \frac{\gamma_\ell^+-\gamma_\ell^-}{\gamma_\ell^+ + \gamma_\ell^-}.
\end{align}
The steady state is thus fully depolarized in the spin eigenbasis and the population imbalance of such eigenstates is ruled by 
\begin{equation}
\langle \hat{\sigma}_\ell^z \rangle_{\sigma} = \frac{\gamma_\ell^+-\gamma_\ell^-}{\gamma_\ell^+ + \gamma_\ell^-} = \frac{\sqrt{1+r_\ell^2}}{1 + r_\ell^2/2} \ .
\end{equation}
From Eq.~\eqref{eq:Bloch} and the quantum regression theorem~\cite{Carmichael1999, Breuer2007}, the relevant correlation functions are determined by
\begin{equation}
\frac{d}{dt}\langle [\sigma_\ell^\pm(t)]^\dagger \sigma_\ell^\pm(0)\rangle = (\pm i\mathcal{E}_\ell - \Gamma_\ell) \langle [\sigma_\ell^\pm(t)]^\dagger \sigma_\ell^\pm(0)\rangle,
\end{equation}
and in the steady state they thus read
\begin{align}
 \langle \hat{\sigma}_\ell^+(t) \hat{\sigma}_\ell^-(0) \rangle_{\sigma} =& e^{-(\Gamma_\ell +  i  \mathcal{E}_\ell) t} \langle \hat{\sigma}_\ell^+(0) \hat{\sigma}_\ell^-(0) \rangle_{\sigma}, \nonumber \\
 = & \frac{\gamma_\ell^+}{\gamma_\ell^+ + \gamma_\ell^-}e^{-(\Gamma_\ell +  i \mathcal{E}_\ell) t} ,\\
 \langle \hat{\sigma}_\ell^-(t) \hat{\sigma}_\ell^+(0) \rangle_{\sigma} =&  \frac{\gamma_\ell^-}{\gamma_\ell^+ + \gamma_\ell^-}e^{-(\Gamma_\ell -  i \mathcal{E}_\ell) t}.
\end{align}
Inserting into Eq.~\eqref{eq:spectrum} gives the Lorentzian quantum noise spectrum
\begin{equation} \label{eq:spectrum_pm}
\mathcal{S}_\ell^\pm(\omega) = \frac{\gamma_\ell^\mp }{\gamma_\ell^+ + \gamma_\ell^-}\frac{2 \Gamma_\ell}{(\omega \pm \mathcal{E}_\ell)^2 + \Gamma_\ell^2}.
\end{equation}
From Eq.~\eqref{eq:spectrum_pm}, one finds that $\mathcal{S}_\ell^+(-\omega) = (\gamma_\ell^-/\gamma_\ell^+) \mathcal{S}_\ell^-(\omega) $, which indicates, together with Eq.~\eqref{eq:rates_posdel} that the ratio $\gamma_\ell^-/\gamma_\ell^+$ controls the ratio between $j\rightarrow i$ and $j\leftarrow i$ transition rates. For a positive energy gap $E_{ij}>0$, the rates of Eq.~\eqref{eq:rates_posdel} can thus be rewritten, using Eq.~\eqref{eq:spectrum_pm}, as 
\begin{subequations}
\begin{align}
& R_{ij}^{(\ell)} = |\chi_{ij}^{(\ell)}|^2 [\gamma_\ell^- \Theta(\delta_\ell) + \gamma_\ell^+ \Theta(-\delta_\ell)] \mathcal{S}_\ell(E_{ij}),  \label{eq:Rijfin} \\
&R_{ji}^{(\ell)} =  |\chi_{ij}^{(\ell)}|^2 [\gamma_\ell^+ \Theta(\delta_\ell) + \gamma_\ell^- \Theta(-\delta_\ell)] \mathcal{S}_\ell(E_{ij}),  \label{eq:Rjifin}
\end{align}
\end{subequations}
where $\Theta(\cdot)$ is Heaviside's Theta function and 
\begin{equation}
\mathcal{S}_\ell(\omega) = \frac{2\Gamma_\ell}{(\omega - |\mathcal{E}_\ell|)^2 + \Gamma_\ell^2}.
\end{equation}

In the main text [Fig.~2(d)], it is shown that the expressions~\eqref{eq:Rijfin} and \eqref{eq:Rjifin} for the rates (obtained tracing out the spins from the model) correctly reproduce the dynamics given by the master equation~\eqref{eq:resspinme} (which, in contrast, includes the spins) by comparing the single-particle evolution obtained using the two models for the parameter values corresponding to the engineered three-state selection pattern of Fig.~2. 

\begin{figure}[t]
\centering
\includegraphics[width=\linewidth]{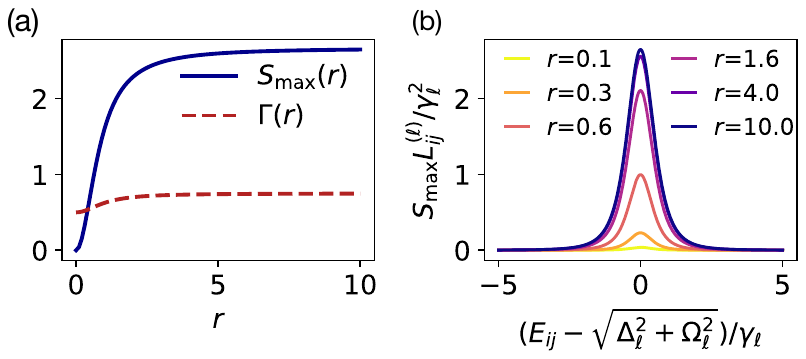}
\caption{(a) Function $\mathcal{S}_{\mathrm{max}}(r)$ and linewidth $\Gamma_\ell(r)/\gamma_\ell$ of Eqs.~\eqref{eq:Smax} and~\eqref{eq:Gammaell} as a function of $r$. (b) Envelope $\mathcal{S}_{\mathrm{max}}(r)L^{(\ell)}_{ij}(r)$ of $A_{ij}^{(\ell)}$ [Eq.~\eqref{eq:Aijl1}, ~\eqref{eq:Smax} and~\eqref{eq:Lij}] for different values of $r$.}
\label{fig:lorentz}
\end{figure}

The asymmetry matrix $A_{ij}^{(\ell)} =R_{ij}^{(\ell)}  - R_{ji}^{(\ell)}$ produced by the rates~\eqref{eq:Rijfin} and \eqref{eq:Rjifin} can be written in the form (for $i>j$)
\begin{align} \label{eq:Aijl1}
A_{ij}^{(\ell)} = \chi_\ell^2 [1-2\Theta(\delta_\ell) ] |M_{ij}^{(\ell)}|^2 \mathcal{S}_{\mathrm{max}}(r_\ell)  L_{ij}^{(\ell)}(r_{\ell}),
\end{align}
where we have introduced the function $\mathcal{S}_{\rm max}(r)$ describing the peak value as a function of $r$, and the function $0\le L_{ij}^{(\ell)}(r) \le 1$ containing the normalized Lorentzian enveloped. These functions explicitly read
\begin{align}
 & \mathcal{S}_{\mathrm{max}}(r) = \frac{8 r^2 }{2 + 3 r^2}, \label{eq:Smax} \\
& L_{ij}^{(\ell)}(r) = \frac{\Gamma_\ell^2(r)}{[E_{ij}-|\delta_\ell|(1+r^2)]^2 + \Gamma_\ell^2(r)} . \label{eq:Lij}
\end{align}
The functions $\mathcal{S}_{\mathrm{max}}(r) $, $\Gamma_\ell(r)$ and the product $\mathcal{S}_{\mathrm{max}}(r) L_{ij}^{(\ell)}(r)$ are shown in Fig.~\ref{fig:lorentz}.

In order to control the system, the asymmetry matrix of Eq.~\eqref{eq:Aijl1} can be decomposed in the form $z_\ell A^{(\ell)}$, with a controllable prefactor $z_\ell$, in different ways:
\begin{enumerate}
\item One possible choice, adopted in the main text, is to assume a tuneable value of the dispersive coupling $\chi_\ell$ up to a maximal value $\chi_{\rm max}$, which can be achieved simply by varying the detuning of the spin's transition frequency from the resonator chain, and a fixed ratio $r_\ell$ for the driving parameters on the spin. In particular, the choice of decomposition is
\begin{align}
z_\ell & = \chi_\ell^2 [1-2\Theta(\delta_\ell)]/\chi^2_{\rm{max}},  \\
A_{ij}^{(\ell)} & = \chi_{\rm{max}}^2|M_{ij}^{(\ell)}|^2 S_{\mathrm{max}}(r_\ell)  L^{(\ell)}_{ij}(r_\ell),
\end{align}
while $\chi_\ell/\chi_{\rm{max}}$ determines the magnitude of the control parameter, $|z_\ell| = \chi_\ell^2/\chi^2_{\rm{max}}$. The sign of $z_\ell$ is given by the sign of the detuning $\delta_\ell$.
\item For fixed resonator-spin couplings $\chi_\ell$, one can adapt the ratios $r_\ell$ between the driving parameters on the spins. Indeed, from Fig.~\eqref{fig:lorentz} one observes that the variation of the spectral linewidth $\Gamma_\ell(r)$ is small, as compared to the variation of the peak value $\mathcal{S}_{\mathrm{max}}(r)$. We may thus approximately treat $\Gamma_\ell(r)$ as constant and vary $r$ as control parameter, yielding the decomposition
\begin{align}
z_\ell = \mathcal{S}_{\mathrm{max}}(r_\ell)  [1-2\Theta(\delta_\ell) ],\\
A_{ij}^{(\ell)} = \chi_\ell^2 |M_{ij}^{(\ell)}|^2  L_{ij}^{(\ell)}(r_\ell).
\end{align}
\end{enumerate}

\section{Values of the simulation parameters and further examples}
\label{sec:simparams}

In this section, the values of the parameters used in the numerical simulations producing the results shown in Figs.~2 and 3 are discussed and used as an example to substantiate the experimental feasibility of the proposed setup in state-of-the-art circuit QED platforms.

\begin{table}[t]
\caption{\label{tab:1}%
Values of $\chi_\ell$, in units of $\chi_{\mathrm{max}}$ and rounded to two significant digits, for the Bose-selection patterns in the five-site chain shown in Fig.~2.
}
\begin{ruledtabular}
\begin{tabular}{llllll}
$S$ & $\chi_0$ & $\chi_1$ & $\chi_2$ & $\chi_3$ & $\chi_4$ \\
\colrule 
\{0, 2, 3\} &  1.0 & 0.92 & 0.79 & 
 0.98 & 0.85 \\
\{1\}& 0.31 & 1.0 & 0.0047 & 0.71 & 0.91 \\
\{2\}& 0.46 & 0.014 & 0.15 & 0.048 & 1.0 
\end{tabular}
\end{ruledtabular}
\end{table}

 For the resonator frequencies $\omega_\ell$, the exact value is not crucial but what is important is that (i) they differ by an amount of the order of the tunnelling coupling $J$, such that the eigenstates are delocalized over several resonators; (ii) accidentally identical gaps in the spectrum are avoided. Similar tight-binding arrays have been realized (also with additional nonlinearities to implement photon-photon interactions, which are not needed here), for instance, in Refs.~\cite{Hacohen2015, Roushan2017, Kollar2019, Yan2019, Ma2019}. We choose arbitrarily $\omega_\ell = \omega_R +  J(\ell/10+ \varepsilon_\ell)$, relative to a common frequency $\omega_R$, where $ \varepsilon_\ell \in [0,1)$ is a disorder parameter picked from one realization of uniformly-distributed random numbers, specifically 
 \begin{multline}
 \bm{\varepsilon}=\big[0.55, 0.73, 0.60, 0.54, 0.42, 0.65, \\
  0.44, 0.89, 0.96,
 0.38\big],
 \end{multline}
 rounded to two significant digits. Although our simulations only depend on the ratio $\omega_\ell/J$, realistic values of $\omega_R$ and $J$ may be chosen, for instance, as $\omega_R/2\pi=5$ GHz and $J/2\pi=30$ MHz~\cite{Blais2021}. The resulting spectrum for the five-site chain is shown in Fig.~2(a). All reservoir spins decay strongly with rate $\gamma_\ell = 0.2 J$. For the choice of $J/2\pi=30$ MHz, this gives $\gamma_\ell/2\pi = 6$ MHz, compatible with values implemented for reservoir transmons, e.g., in Ref.~\cite{Ma2019}. The parameter $\delta_\ell$ is fixed by (i) enforcing that the $\ell$th-spin level splitting $\mathcal{E}_\ell = |\delta_\ell|\sqrt{1+r_\ell^2}$ resonates with a chosen system level distance $E_{ij}$ of the bosonic chain, and (ii) imposing the sign $\mathrm{sgn}(\delta_\ell)$ found from the solution $\bm{z}$ of the inequalities of Eq.~(7) for a given target condensation pattern. The values of the parameter $r_\ell$ are then chosen such that the Rabi frequency $\delta_\ell r_\ell$ of the microwave drives on the qubits meets experimental constraints---in particular, such that it remains around an order of magnitude smaller than the transmon nonlinearities, which have typical values of $\sim 300$ MHz, to prevent unwanted excitations outside of the lowest two-level subspace. For the five-site chain with connectivity of Fig.~2(a), up to the sign of $\delta_\ell$, which depends on the choice of condensation patterns, the values used are then (up to three significant digits)
\begin{align}
 |\delta_\ell|/J & = [0.578, 2.37, 0.989, 0.456, 0.616],\nonumber \\
 r_\ell/J & =  [1.36, 0.552, 1.36, 1.36,1.36 ].
\end{align}
This yields detunings in the range $|\delta_\ell|/2\pi\sim 14-70$ MHz and Rabi frequencies in the range $|\delta_\ell| r_\ell /2\pi\sim 19-40$ MHz for the microwave drives on the transmons as routinely employed. 
The values of the dispersive couplings $\chi_\ell$, which are the physical control parameters leveraging the contribution of the different artificial baths to the total rate-asymmetry matrix, are given by $\chi_\ell=\sqrt{|z_\ell|}\chi_{\rm max}$ from the solution $\bm{z}$ of the Eq.~(7). As discussed in the main text, the tuneability of $\chi_{\ell}$ to these desired values can be attained \textit{in situ} in different ways, depending on the specific choice of superconducting-qubit architecture. Since the dispersive coupling descends from a Jaynes-Cummings-type resonator-spin coupling (of strength $g$) in the limit in which the qubit and the resonator have a large frequency mismatch $\Delta$, according to $\chi_\ell \sim g^2/\Delta$, possible options are as follows 
\begin{itemize}
\item[(i)] for tuneable-frequency transmons, adapting $\Delta$ by biasing the transmon's frequency~\cite{Krantz2019};
\item[(ii)] for tuneable-coupling qubits, varying $g$~\cite{Chen2014, Roushan2017b, Yan2018};
\item[(iii)] a combination of both (i) and (ii) for architectures supporting both frequency and coupling tuning as used, e.g., in \cite{Arute2019}.
\end{itemize}
We chose $\chi_{\rm max}/J=0.15$, corresponding, for instance, to $g/2\pi=50$ MHz and $\Delta/2\pi=300$ MHz. For the single-state condensation of Fig.~2(c) and the three-state selection of Fig.~2(b) and~2(d)-(f), the values of $|z_\ell|$ are given in Table~\ref{tab:1}.

\begin{figure}
\centering
\includegraphics[width=0.8\linewidth]{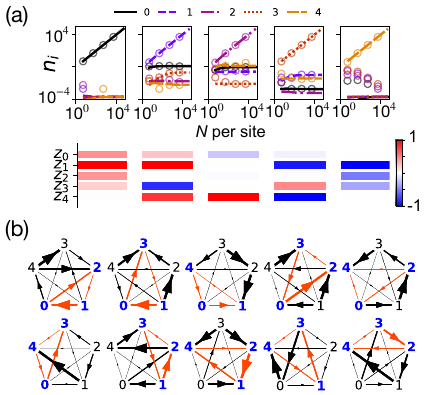}
\caption{Additional results for the five-site system of Fig.~2. (a) Results for controlled condensation in any choice of single state, and corresponding values of the control variables. (b) Asymmetry-matrix networks corresponding to engineered three-state selection in any possible triplet of states (blue-coloured).}
\label{fig:from_main}
\end{figure}

For similar values, protocols are found for controlled condensation in any single state, as shown in Fig.~\ref{fig:from_main}(a). Also, selection in any triplet of states is possible: the asymmetry-matrix networks found for each target three-state set $\{\ket{i}, \ket{j>i}, \ket{k>j}\}$, with the same occupation fractions of the example in the main text ($n_i=1/10$, $n_j=3/10$, $n_k=6/10$) are depicted in Fig.~\ref{fig:from_main}(b).

For the ten-site chain of Fig.~3, instead, the values of $\delta_\ell$ used are
\begin{align}
|\delta_0| =& 2 E_{21}/3 \simeq 0.288 J,  & |\delta_5|  = 2 E_{75}/3 \simeq 0.671 J, \nonumber \\
|\delta_1| =& 2 E_{70}/3 \simeq  2.24 J, & |\delta_6| = 2 E_{42}/3 \simeq 0.591 J, \nonumber \\
|\delta_2| =& 2 E_{74}/3 \simeq 1.10 J, & |\delta_7| = 2 E_{95}/3 \simeq 1.27 J, \nonumber \\
|\delta_3| =& 2 E_{52}/3 \simeq 1.02 J,  & |\delta_8| = 2 E_{97}/3 \simeq 0.599 J ,\nonumber \\
|\delta_4| =& 2 E_{84}/3 \simeq 1.34 J , &|\delta_9| = 2 E_{43}/3 \simeq 0.287 J. 
\end{align}
and $r_\ell\simeq \sqrt{5}\delta_\ell/3$. The values of $\chi_\ell$ found through Eq.~(7), giving the requested equal occupancy of the selected states, are given in Table~\ref{tab:2}.

\begin{table}
\caption{Values of $\chi_\ell$, in units of $\chi_{\rm max}=0.15J$, corresponding to the ten-site selection pattern shown in Fig.~3.
\label{tab:2}}
\begin{ruledtabular}
\begin{tabular}{llllllllll}
 $\chi_0$ & $\chi_1$ & $\chi_2$ & $\chi_3$ & $\chi_4$ & $\chi_5$ & $\chi_6$ & $\chi_7$ & $\chi_8$ & $\chi_9$\\
 \colrule
 0.94 & 
 0.24 &
0.79 &
 1.0 &
 1.0 &
 0.93 &
0.85 &
 0.91 &
 1.0& 0.24 
\end{tabular}
\end{ruledtabular}
\end{table}

\section{Further experimental considerations}
 In this Section, we discuss further aspects of relevance for potential experimental implementations.\\
 
\textbf{Temperature effects and photon emission.} At the dilution-refrigerator temperatures $\sim 10$ mK at which superconducting circuits operate, the absorption of thermal photons is strongly suppressed and would only contribute a small addition of particles to the Bose selection. Considering superconducting microwave resonators with routinely-achieved $Q$-factors of $Q\sim 10^5-10^6$, the photon decay rate $\kappa/2\pi\sim 1-10$ kHz by spontaneous emission are one to two orders of magnitude smaller than the dissipation engineered through the reservoir spins. The same applies to weak losses potentially originating from Purcell-like decay due to the weak hybridization of the resonators with the strongly damped reservoir spins, which can be estimated as $(g_\ell/\Delta_\ell)^2 \gamma_\ell$~\cite{Blais2021} (with $g$ and $\Delta$ defined in Sec.~\ref{sec:simparams} and $\gamma_\ell$ the spin decay rate). The net particle loss will thus be weak within the timescales necessary for the system to reach its engineered Bose-selected steady state. Also, as discussed in the main text and in detail in Sec.~\ref{app:pumploss}, Bose selection is driven by terms in the mean-field Eq.~\eqref{eq:meanfield1}, which are quadratic in the large occupations $n_i$, whereas losses by photon emission depend linearly on $n_i$. Therefore, `pre-relaxation' to the Bose selected state occurs much faster than the photon leakage. Finally, the system can be fully stabilized also against this weak photon loss, such that losses are sufficiently compensated and do not alter the Bose selection pattern also at long times, by using additional qubit reservoirs as particle pumps. This is detailed below, in Sec.~\ref{app:pumploss}.

\textbf{Interactions.} The perfectly harmonic spectrum of the microwave resonators provides a system hosting non-interacting photons, as desired to study Bose selection driven purely by quantum statistics. Spurious photon-photon interactions may in principle arise due to hybridization with the ancillary reservoir spins, inducing effective nonlinearities in the resonator array. However, since the spins operate in the dispersive regime, the latter (as well as resonant photon exchange yielding particle loss) can be neglected. This is, in a way, substantiated by the fact that tracing out the spins in favour of effective quantum-jump rates $R_{ij}$ among the eigenstates of the resonators system yields a very good description, as shown quantitatively in Fig.~2(d). Spin-spin interaction mediated by the resonators, potentially influencing the engineered-dissipation, may arise in case the qubits happen to have the same transition frequency~\cite{Majer2007}, but this will in general require fine-tuning, and thus also this effect is not expected to play a significant role in the proposed implementation. \\

\begin{figure}
\includegraphics[width=0.8\linewidth]{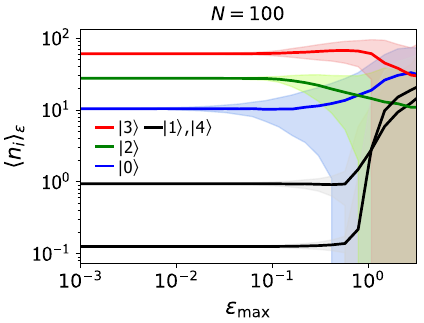}
\caption{Robustness of the engineered condensation pattern of Fig.~2 with respect to perturbations of the control parameters $\chi_j$ (see the discussion in the main text and Eq.~\eqref{eq:perturb}). The shaded areas around the curves indicate the region within one standard deviation from the average value $\langle n_i\rangle_\epsilon$. }
\label{fig:robustness}
\end{figure}
\textbf{Robustness of the control parameters.} The engineered Bose selection patterns are robust against imperfections in the values of the control parameters, unless the system is engineered on purpose in the vicinity of a transition point towards other BS patterns as done for the BEC switch in the main text. To show this, we analyze the stability of the three-state BS pattern of Fig.~2 in Fig.~\ref{fig:robustness}. We study the variation of a snapshot of the system state for $N=100$ particles at time $Jt=200$, where the steady state in perfect control conditions is basically attained [see Fig.~2(f)], in response to a perturbation of the ideal control values of the form
\begin{equation} \label{eq:perturb}
\chi_\ell^2 \to \chi_\ell^2 (1 + \epsilon_\ell),
\end{equation}
where $\epsilon_\ell$ are uniformly distributed random numbers in an interval $[-\epsilon_{\rm max}, \epsilon_{\rm max}]$. The relative occupations of the single-particle eigenstates is shown in Fig.~\ref{fig:robustness} as a function of $\epsilon_{\rm max}$, where each point is the average $\langle n_i\rangle_\epsilon$ over $10^3$ realizations of sets of $\epsilon_j$ bounded by the corresponding $\epsilon_{\rm max}$. The shaded area around the curves indicates the region within one standard deviation $\sigma_i=\sqrt{\langle n_i^2\rangle_\epsilon-\langle n_i\rangle_\epsilon^2}$ from the average value. The occupations start to deviate significantly from the target values only are for $\epsilon_{\rm max}$ beyond 10\% and the Bose-selected state is disrupted for $\epsilon_{\rm max}\sim100$\%, as signalled by the depletion of the selected states and the development, on average, of a macroscopic occupation also of other states.\\

\textbf{Finite qubit nonlinearities.} In the circuit-QED implementation proposed, the ancillary reservoir spins are implemented, for instance, via transmon qubits (as done, e.g., in Ref.~\cite{Ma2019}). The latter possess a multilevel spectrum with a strong nonlinearity which enables one to treat them effectively as two-level systems (thus approximating the nonlinearity as infinite). A \textit{finite} nonlinearity can be predicted to impact the results presented here predominantly with two effects. (i) The actual dispersive coupling will tend to be a bit smaller than the prediction $g^2/\Delta$, yielding slightly smaller engineered dissipation rates. Nonetheless, this can be taken into account when tuning the qubit frequency or the resonator-qubit coupling to control $\chi_\ell$ (see Sec.~\ref{sec:simparams}), in order to attain the values prescribed by the control algorithm for realizing a target condensation pattern. (ii) The slightly off-resonant microwave drive on the qubit transition may induce off-resonant excitation outside of the qubit subspace to higher excited states. The primary potential excitation channel is given by accidental two-photon transitions to the doubly excited state. However, since the values of the drive detuning $\delta_\ell$ considered in Sec.~\eqref{sec:simparams} are well away from the two-photon resonance, these processes will be suppressed and will not represent an important limitation. Moreover, the driving amplitudes used (see Sec.~\ref{sec:simparams}) are maintained around an-order of magnitude smaller than the nonlinearity (as commonly used) to suppress unwanted excitations.\\

\textbf{Intial particle loading.} The system can be initialized with a desired number of photons by injecting them through the reservoir spins using different approaches. One method is to use the particle pumps proposed in Sec.~\eqref{app:pumploss}. Alternatively, the strategy of Ref.~\cite{Hofheinz2008} can be used, in which a reservoir qubit is periodically excited and brought in and out of resonance with a resonator by rapidly tuning its transition frequency: on resonance, the excitation tunnels to the resonator with a half Rabi rotation; then, the qubit is effectively decoupled by biasing its transition frequency away from the resonator and is then re-excited for the next injection cycle. A third method was implemented in Ref.~\cite{Ma2019}, which can be intuitively summarized as follows. A reservoir transmon (with eigenstates $\ket{n}$) is tuned such that its $\ket{1}\leftrightarrow\ket{2}$ transition is in resonance with the resonator, i.e. has the energy of a resonator photon. One photon injection is achieved by first exciting the transmon to its second-excited state $\ket{2}$ via a weak two-photon drive. The resonator, strongly coupled to the transmon, can then resonantly absorbs one photon from it, leaving it in state $\ket{1}$. A strong $\ket{1}\to\ket{0}$ transmon decay quickly relaxes it to its ground state $\ket{0}$, preventing it from re-absorbing the photon, since such a process is then off-resonant. The net result is then a photon injection in the resonator system.

\section{Impact of pump and loss} \label{app:pumploss}

We here discuss how particle pumping and loss from the resonators contribute to the mean-field equations, how they impact the Bose selection patterns, and how a pump for the resonators can be implemented.
\begin{figure}[t]
\centering
\includegraphics[width=\linewidth]{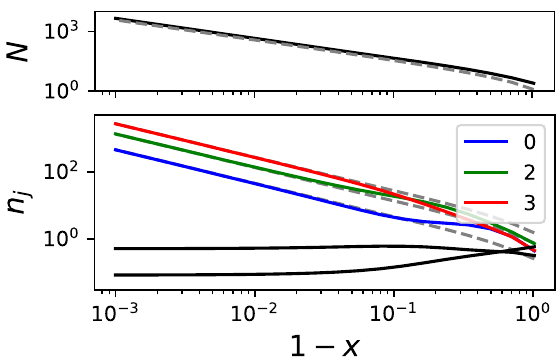}
\caption{Bose selection in the steady state of the five-site chain with pump and loss, as a function of the imbalance between pump and loss rates (lower panel). The latter are chosen as $K_i^-=R_i\times 0.1J$ and $K_i^+=(1-x) K_i^-$ where $R_i, R_i'$ are random numbers distributed uniformly in $[0,1)$. The mean steady-state total number of particles $N$ is depicted in the upper panel (solid line), where the dashed line shows the analytically predicted bound of Eq.~\eqref{eq:Nestimate}.}
\label{fig:pumploss}
\end{figure}
\begin{figure}
\centering
\includegraphics[width=\linewidth]{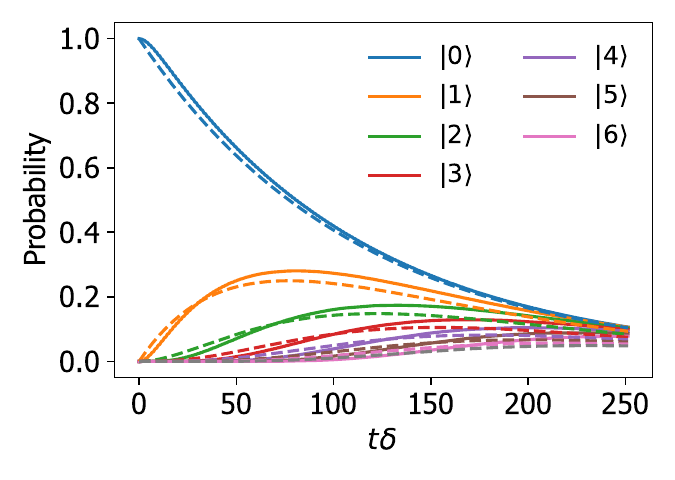}
\caption{Probability of being in the different Fock states of the oscillator, with spin and oscillator initialized in their ground state. Solid lines are obtained by solving the master equation~\eqref{eq:me_pump}, while dotted lines are obtained by the dissipator $\kappa^+ D[\hat a^\dagger]\rho$ with the predicted effective rate $\kappa^+ = g^2/\gamma$. The parameter values used are $g=0.06\delta$, $\gamma=0.4\delta$. }
\label{fig:ppump}
\end{figure}

In the presence of pumping and decay for each resonator, the master equation for the resonator chain reads
\begin{equation} \label{eq:ploss}
\frac{d\hat{\rho}}{dt} = -i[\hat{H}_S, \hat{\rho}] + \sum_{\ell=1}^M \big(\kappa^+_\ell D[\hat{a}^\dagger_\ell]\hat{\rho} + \kappa_\ell^- D[\hat{a}_\ell]\hat{\rho}\big),
\end{equation}
where $\hat{H}_S$ is the chain Hamiltonian of Eq.~\eqref{eq:resspinme} and $\kappa_\ell^\pm$ are the pump ($+$) and loss ($-$) rates for the $j$th resonator. For clarity, in Eq.~\eqref{eq:ploss}, we do not include the particle-number-conserving coupling to the artificial baths, whose contribution to the mean field equations can be treated separately, giving the mean-field Eq.~\eqref{eq:meanfield1}. Formally diagonalizing the Hamiltonian $\hat{H}_S = \sum_j E_j \hat{c}_j^\dagger \hat{c}_j$ through a transformation $\hat{c}_j = \sum_\ell V_{j\ell} \hat{a}_\ell$, the master equation in the chain eigenbasis reads
\begin{multline} \label{eq:stepx}
\frac{d\hat{\rho}}{dt} = -i[\hat{H}_S,\hat{\rho}] + \sum_{\ell=1}^M \Big(\kappa^+_\ell D\big[\sum_i V_{i\ell} \hat{c}_i^\dagger\big]\hat{\rho} \\
+ \kappa_\ell^- D \big[\sum_i V_{i\ell}^* \hat{c}_i \big]\hat{\rho}\Big).
\end{multline}
Adopting the secular approximation~\cite{Breuer2007}, valid if the decay rates $\kappa_\ell^\pm$ are much smaller than the gaps in the system and their difference, Eq.~\eqref{eq:stepx} reduces to
\begin{equation}
\frac{d\hat{\rho}}{dt} = -i[\hat{H}_S, \hat{\rho}] + \sum_i (K_i^+  D[\hat{c}_i^\dagger]\hat{\rho} +  K_i^- D[\hat{c}_i]\hat{\rho}),
\end{equation}
with modified decay rates $K_i^\pm = \sum_{\ell=1}^M \kappa^\pm_\ell |V_{i\ell}|^2.$ Considering the mean occupation $n_j(t) = \mathrm{tr}[\hat{\rho}(t) \hat{n}_j] $ of the $j$th eigenstate, this yields in turn
\begin{equation} \label{eq:pumpterm}
\frac{dn_j(t)}{dt} = K_j^+ (n_j+1) - K_j^- n_j.
\end{equation}
The terms on the r.h.s. of Eq.~\eqref{eq:pumpterm} are then the pump-loss contribution to the mean-field equations to be included in Eq.~\eqref{eq:meanfield1}. They are linear in the $n_j$, in contrast to the terms driving Bose selection, which are quadratic. From Eq.~\eqref{eq:pumpterm}, the total steady-state particle number $N=\sum_j n_j$ in the system can be estimated by summing over $j$,
\begin{align}
0 & =  \sum_j (K_j^+ - K_j^-) n_j  + \sum_j K_j^{+},\\
 & < N \mathrm{max}_j (K_j^+ - K_j^-) + \sum_j K_j^{+},
\end{align}
which gives the estimate
\begin{equation} \label{eq:Nestimate}
N > \frac{ \sum_j K_j^{+}}{\mathrm{max}_j (K_j^- - K_j^+)}\ .
\end{equation}
 We analyze in Fig.~\ref{fig:pumploss} the fate of the Bose selection pattern obtained in the steady state of the five-site chain of Fig.~2(c) when subject to pumping and loss with rates ratio $K_i^+/K_i^-=1-x$ for variable $x$. From the lower panel of Fig.~\ref{fig:pumploss}, one can see that nonequilibrium Bose condensation as predicted is indeed quickly attained as the pump rates approach the loss rates, and that the total mean number of particles increases accordingly [upper panel of Fig.~\ref{fig:pumploss}]. The set of selected states is the one dictated by the controlled asymmetry matrix $A$, as predicted. Selection in the desired states is immediately attained also for weak pumping, and the chosen relative occupations are also recovered when the pump rates reach about 90$\%$ of the loss rates, confirming that Bose selection can successfully be controlled also for a leaky system. 

We next discuss how to engineer a pump for the resonators, namely the terms $\kappa^+_\ell D[a^\dagger_\ell]\rho$ of Eq.~\eqref{eq:ploss}. We consider an auxiliary spin coupled in resonance to a resonator with a tuneable coupling $g(t)$. The Hamiltonian reads
\begin{equation}
\hat{H}(t) = \frac{\delta}{2} \hat{\sigma}_z + g(t)(\hat{\sigma}_+  + \hat{\sigma}_-) (\hat{a} + \hat{a}^\dagger)+ \delta \hat{a}^\dagger \hat{a}\ ,
\end{equation}
with $\delta\gg g(t)$. The spin is further strongly decaying with rate $g(t)\ll \gamma \ll \delta$, such that the qubit-oscillator master equation reads
\begin{equation} \label{eq:me_pump}
\frac{d\hat{\rho}}{dt} = -i [\hat{H}(t), \hat{\rho}] + \gamma D[\hat{\sigma}_-] \hat{\rho}\ .
\end{equation}
We consider a drive of the coupling, of the form $g(t) = g\cos(2\delta t)$, which makes resonant counter-rotating terms $\hat{a}^\dagger \hat{\sigma}_+$ and $\hat{a} \hat{\sigma}_-$, such that, in interaction picture with respect to the bare Hamiltonians and in rotating wave approximation, the Hamiltonian becomes
\begin{equation}
H =  \frac{g}{2}(\hat{\sigma}_+ \hat{a}^\dagger  + \hat{\sigma}_- \hat{a} ) \ .
\end{equation}
Because of the strong damping, only processes induced by $\hat{a}^\dagger \hat{\sigma}_+$ can effectively take place, followed by a quick resetting of the spin to the de-excited state, thus pumping the resonator. 
We can apply the results of Sec.~\ref{app:spinreservoirs}, in particular following the derivations starting from Eq.~\eqref{eq:anontherH} with, in this case, $\delta_\ell=0$. The transition rates between two oscillator states $\ket{n}$ and $\ket{n+1}$ read
\begin{align}
R_{n+1,n} & = \left(\frac{g}{2}\right)^2 |\!\bra{n+1} \hat{a}^\dagger \ket{n}\!|^2 S(0)\ ,\\
& = g^2 (n+1) \mathcal{S}(0)/4 \ ,
\end{align}
where $\mathcal{S}(\omega)$ is the spectral density
\begin{equation}
\mathcal{S}(\omega) = \frac{\gamma}{\omega ^2 +(\gamma/2)^2}\ .
\end{equation}
Due to the harmonicity of the oscillator, the spin drives all $\ket{n}\to\ket{n+1}$ transitions simultaneously, realizing the desired pump with a resonant pump rate $\kappa^+ = g^2/\gamma$. 
The achievement of a pump as predicted is verified in Fig.~\ref{fig:ppump}, where the oscillator evolution produced by the master Eq.~\eqref{eq:me_pump} is compared with that given by the effective pump master equation $d\hat{\rho}/dt = \omega 
\hat{a}^\dagger \hat{a} + \kappa^+ D[\hat{a}^\dagger]$.
\end{document}